\documentclass[12pt,oneside,reqno,titlepage, hyperfootnotes=false]{amsart}
\usepackage{lmodern}
\usepackage[T1]{fontenc}
\usepackage[latin9]{inputenc}
\usepackage[a4paper]{geometry}
\usepackage{color}
\usepackage{array}
\usepackage{bm}
\usepackage{amstext}
\usepackage{amsthm}
\usepackage{amssymb}
\usepackage{graphicx}
\usepackage{setspace}
\usepackage[authoryear]{natbib}
\usepackage[unicode=true,pdfusetitle,
 bookmarks=true,bookmarksnumbered=false,bookmarksopen=false,
 breaklinks=false,pdfborder={0 0 0},pdfborderstyle={},backref=false,colorlinks=true]
 {hyperref}

\makeatletter

\providecommand{\tabularnewline}{\\}

\numberwithin{equation}{section}
\numberwithin{figure}{section}

\usepackage{amsfonts}
\usepackage{amstext}
\usepackage{amsthm}

\usepackage{hyperref}
\hypersetup{
    colorlinks = true,
    citecolor = black
}

\usepackage{threeparttable}
\usepackage{graphicx}
\usepackage{enumerate}
\usepackage{listings}
\usepackage{subcaption}

\DeclareMathOperator*{\argmin}{arg\,min}

\DeclareMathOperator*{\argmax}{arg\,max}

\usepackage{float}
\newtheorem{thm}{Theorem}

\newtheorem{lem}{Lemma}
\newtheorem*{lem*}{Lemma}
\newtheorem*{thm*}{Theorem}
\newtheorem{cor}{Corollary}
\newtheorem*{asm1}{Assumption 1}

\theoremstyle{definition}

\makeatother

\begin{document}
\title{How to sample and when to stop sampling: The generalized Wald problem
and minimax policies}
\author{Karun Adusumilli$^\dagger$}
\begin{abstract}
We study sequential experiments where sampling is costly and a decision-maker
aims to determine the best treatment for full scale implementation
by (1) adaptively allocating units between two possible treatments,
and (2) stopping the experiment when the expected welfare (inclusive
of sampling costs) from implementing the chosen treatment is maximized.
Working under a continuous time limit, we characterize the optimal
policies under the minimax regret criterion. We show that the same
policies also remain optimal under both parametric and non-parametric
outcome distributions in an asymptotic regime where sampling costs
approach zero. The minimax optimal sampling rule is just the Neyman
allocation: it is independent of sampling costs and does not adapt
to observed outcomes. The decision-maker halts sampling when the product
of the average treatment difference and the number of observations
surpasses a specific threshold. The results derived also apply to
the so-called best-arm identification problem, where the number of
observations is exogenously specified.\textcolor{blue}{{} }
\end{abstract}

\thanks{\textit{This version}: \today{}\\
The paper subsumes an unpublished note previously circulated as ``Neyman
allocation is minimax optimal for best arm identification with two
arms'' on ArXiv at the following link: \href{https://arxiv.org/abs/2204.05527}{https://arxiv.org/abs/2204.05527}.
\\
\thispagestyle{empty}I would like to thank the editor and three anonymous
referees for valuable comments that substantially improved the paper.
Thanks also to Tim Armstrong, Federico Bugni, David Childers, Pepe
Montiel-Olea, Chao Qin, Azeem Shaikh, Tim Vogelsang and seminar participants
at various universities and conferences for helpful comments and suggestions.\\
$^\dagger$Department of Economics, University of Pennsylvania}
\maketitle

\section{Introduction \protect\label{sec:Introduction}}

Acquiring information is expensive. Experimenters need to carefully
choose how many units of each treatment to sample and when to stop
sampling. This paper seeks to develop techniques for incorporating
the cost of information into experimental design. Specifically, we
focus our analysis of costly experimentation within the context of
comparative trials where the aim is to determine the best of two treatments. 

In the computer science literature, such experiments are referred
to as A/B tests. Technology companies like Amazon, Google and Microsoft
routinely run hundreds of A/B tests a week to evaluate product changes,
such as a tweak to a website layout or an update to a search algorithm.
However, experimentation is expensive, especially if the changes being
tested are very small and require evaluation on large amounts of data;
e.g., \citet{deng2013improving} state that even hundreds of millions
of users were considered insufficient at Google to detect the treatment
effects they were interested in. Clinical or randomized trials are
another example of A/B tests. Even here, reducing experimentation
costs is a key goal. In fact, this has been a major objective for
the FDA since 2004 when it introduced the `Critical Path Initiative'
for streamlining drug development; this in turn led the FDA to promote
sequential designs in clinical trials (see, e.g., \citealp{FDApolicybrief},
for the current guidance, which was influenced by the need to reduce
experimentation costs). For this reason, many recent clinical trials,
such as the ones used to test the effectiveness of Covid vaccines
(e.g., \citealp{baden2021efficacy}), now use multi-stage designs
where the experiment can be terminated early if a particularly positive
or negative effect is seen in early stages.

In practice, the cost of experimentation directly or indirectly enters
the researchers' experimental design when they choose an implicit
or explicit stopping time (note that we use stopping time interchangeably
with the number of observations in the experiment). For instance,
in testing the efficacy of vaccines, experimenters stop after a pre-determined
number of infections. In other cases, a power analysis may be used
to determine sample size before the start of the experiment. But if
the aim is to maximize social welfare (or profits), neither of these
procedures is optimal.\footnote{See, e.g., \citet{manski2016sufficient} for a critique on the common
use of power analysis for determining the sample size in randomized
control trials. } 

In this paper, we develop optimal experimentation designs that maximize
welfare while also taking into account the cost of information. In
particular, we study optimal sampling and stopping rules in sequential
experiments where sampling is costly and the decision maker (DM) aims
to determine the best of two treatments by: (1) adaptively allocating
units to one of these treatments, and (2) stopping the experiment
when the expected welfare, inclusive of sampling costs, is maximized.
We term this the generalized Wald problem, and use minimax regret
(\citealp{manski2021econometrics}), a natural choice criterion under
ambiguity aversion, to determine the optimal decision rule.\footnote{We do not consider the minimax risk criterion as it leads to a trivial
decision: the DM should never experiment and always apply the status
quo treatment. } 

We first derive the optimal decision rule in continuous time, under
the so-called diffusion regime \citep{fan2021diffusion,wager2021diffusion}
where information arrives gradually in the form of (continuous) Gaussian
increments. Then, we show that analogues of this decision rule are
also asymptotically optimal under parametric and non-parametric distributions
of outcomes. The asymptotics, which appear to be novel, involve taking
the marginal cost of experimentation to $0$ at a specific rate. Section
\ref{sec:Formal-properties} delves into the rationale behind these
`small cost asymptotics', and argues that they are practically quite
relevant. It is important to clarify here that `small costs' need
not literally imply the monetary costs of experimentation are close
to $0$. Rather, it denotes that these costs are small compared to
the benefit of choosing the best treatment for full-scale implementation.

The optimal decision rule has a number of interesting, and perhaps,
surprising properties. First, the optimal sampling rule is history
independent and also independent of sampling costs. In fact, it is
just the Neyman allocation, which is well known in the Randomized
Control Trial (RCT) literature as the (fixed) sampling strategy that
minimizes estimation variance; our results state that one cannot do
better than this even when allowing for adaptive strategies. Second,
it is optimal to stop when the difference in average outcomes between
the treatments, multiplied by the number of observations collected
up to that point, exceeds a specific threshold. The threshold depends
on sampling costs and the standard deviation of the treatment outcomes.
Finally, at the conclusion of the experiment, the DM chooses the treatment
with the highest average outcomes. The decision rule therefore has
a simple form that makes it attractive for applications.

Our results also apply to the best arm identification problem with
two arms.\footnote{The results for best arm identification were previously circulated
in an unpublished note by the author, accessible from ArXiV at \href{https://arxiv.org/abs/2204.05527}{https://arxiv.org/abs/2204.05527}.
The current paper subsumes these results.} Best arm identification shares the same aim of determining the best
treatment but the number of observations is now exogenously specified,
even as the sampling strategy is allowed to be adaptive. Despite this
difference, we find Neyman allocation to be the minimax-regret optimal
sampling rule in this context as well. However, by not stopping adaptively,
we lose on experimentation costs. Compared to best arm identification,
we show that the use of an optimal stopping time allows us to attain
the same regret, exclusive of sampling costs, with $40\%$ fewer observations
on average (under the least-favorable prior); this is independent
of model parameters such as sampling costs and outcome variances. 

For the most part, this paper focuses on constant sampling costs (i.e.,
constant per observation). This has been a standard assumption since
the classic work of \citet{wald1947sequential}, see also \citet{arrow1949bayes}
and \citet{fudenberg2018speed}, among others. In fact, many online
marketplaces for running experiments, e.g., Amazon Mechanical Turk,
charge a fixed cost per query/observation. Note also that the costs
may be indirect: for online platforms like Google or Microsoft that
routinely run thousands of A/B tests, these could correspond to how
much experimentation hurts user experience. Still, one may wonder
whether and how our results change under other cost functions and
modeling choices, e.g., when data is collected in batches, or, when
we measure regret in terms of nonlinear or quantile welfare. We asses
this in Section \ref{sec:Variations-and-extensions}. Almost all our
results still go through under these variations. We also identify
a broader class of cost functions, nesting the constant case, in which
the form of the optimal decision stays the same. 

\subsection{Related literature}

The question of when to stop sampling has a rich history in economics
and statistics. It was first studied by \citet{wald1947sequential}
and \citet{arrow1949bayes} with the goal being hypothesis testing,
specifically, optimizing the trade-off between type I and type II
errors, instead of welfare maximization. Still, one can place these
results into the present framework by imagining that the distributions
of outcomes under both treatments are known, but it is unknown which
distribution corresponds to which treatment. This paper generalizes
these results by allowing the distributions to be unknown. For this
reason, we term the question studied here the generalized Wald problem. 

\citet{chernoff1959sequential} studied the sequential hypothesis
testing problem under multiple hypotheses, using large deviation methods.
The asymptotics there involve taking the sampling costs to 0, even
as there is a fixed reward gap between the treatments. More recently,
the stopping rules of \citet{chernoff1959sequential} were incorporated
into the $\delta$-PAC (Probably Approximately Correct) algorithms
devised by \citet{garivier2016optimal} and \citet{qin2017improving}
for best arm identification with a fixed confidence. The aim in these
studies is to minimize the amount of time needed to attain a pre-specified
probability, $1-\delta$, of selecting the optimal arm. However, these
algorithms do not directly minimize a welfare criterion, and the constraint
of pre-specifying a $\delta$ could be misplaced, if, e.g., there
is very little difference between the first and second best treatments.
In fact, under the least-favorable prior, our minimax decision rule
mis-identifies the best treatment about 23\% of the time. \citet{qin2022adaptivity}
study the costly sampling problem under fixed reward gap asymptotics
using large deviation methods. The present paper differs in using
local asymptotics and in appealing to a minimax regret criterion.
However, unlike the papers cited above, we only study binary treatments.

A number of papers (\citealp{colton1963model,lai1980sequential,chernoff1981sequential})
have studied sequential trials in which there is a population of $N$
units, and at each period, the DM randomly selects two individuals
from this population, and assigns them to the two treatments. The
DM is allowed to stop experimenting at any point and apply a single
treatment on the remainder of the population. The setup in these papers
is intermediate between our own and two-armed bandits: while the aim,
as in here, is to minimize regret, acquiring samples is not by itself
expensive and the outcomes in the experimentation phase matter for
welfare. This literature also does not consider optimal sampling rules. 

The paper is also closely related to the growing literature on information
acquisition and design, see, \citet{hebert2017rational,fudenberg2018speed,morris2019wald,liang2022dynamically},
among others. \citet{fudenberg2018speed} study the question of optimal
stopping when there are two treatments and the goal is to maximize
Bayes welfare (which is equivalent to minimizing Bayes regret) under
normal priors and costly sampling. While their approach relies on
an exogenously specified sampling rule, \citet{liang2022dynamically}
extend this line of inquiry by allowing for endogenous selection of
the sampling rule. In fact, for constant sampling costs, the setup
in \citet{liang2022dynamically} is similar to ours but the welfare
criterion is different: their framework adopts a Bayesian perspective
with normal priors. Although the Neyman allocation plays a key role
in the optimal sampling rules under both frameworks, the optimal stopping
times have very different qualitative and quantitative properties.
A detailed comparison is provided in Section \ref{subsec:Discussion}.
These differences in stopping times arise because the minimax regret
criterion corresponds to a least-favorable prior with a specific two-point
support. Thus, our results highlight the important role played by
the prior in determining even the qualitative properties of optimal
decisions. This motivates the need for robust decision rules, and
the minimax regret criterion is one way to obtain them. 

Our results also speak to the literature on drift-diffusion models
(DDMs), which are widely used in neuroscience and psychology to study
choice processes \citep{luce1986response,ratcliff2008diffusion,fehr2011neuroeconomic}.
DDMs are based on the classic binary state hypothesis testing problem
of \citet{wald1947sequential}. \citet{fudenberg2018speed} extend
this model to allow for continuous states, using Gaussian priors,
and show that the resulting optimal decision rules are very different,
even qualitatively, from the predictions of DDM. In this paper, we
show that if the DM is ambiguity averse and uses the minimax regret
criterion, then the predictions of the DDM model are recovered even
under continuous states. In other words, decision making under ignorance
can bring us back to DDM.

Finally, the results in this paper are unique in regards to all the
above strands of literature in showing that any discrete time parametric
and non-parametric version of the problem can be reduced to the diffusion
limit under small cost asymptotics. Diffusion asymptotics were introduced
by \citet{fan2021diffusion} and \citet{wager2021diffusion} to study
the properties of Thompson sampling in bandit experiments. The techniques
for showing asymptotic equivalence to the limit experiment build on,
and extend, previous work on sequential experiments by \citet{adusumilli2021risk}.
Relative to that paper, the novelty here is two-fold: first, we derive
a sharp characterization of the minimax optimal decision rule for
the Wald problem. Second, we introduce `small cost asymptotics' that
may be of independent interest in other, related problems where there
is a `local-to-zero' cost of continuing an experiment. 

\section{Setup under incremental learning\protect\label{sec:Diffusion-asymptotics-and}}

Following \citet{fudenberg2018speed} and \citet{liang2022dynamically},
we start by describing the problem under a stylized setting where
time is continuous and information arrives gradually in the form of
Gaussian increments. In statistics and econometrics, this framework
is also known as diffusion asymptotics \citep{adusumilli2021risk,fan2021diffusion,wager2021diffusion}.
The benefit of the continuous time analysis is that it enables us
to provide a sharp characterization of the minimax optimal decision
rule; this is otherwise obscured by the discrete nature of the observations
in a standard analysis. Section \ref{sec:Formal-properties} describes
how these asymptotics naturally arise under a limit of experiments
perspective when we employ small-cost asymptotics and a local-to-zero
scaling for the treatment effect. 

The setup is as follows. There are two treatments $0,1$ corresponding
to unknown mean rewards $\bm{\mu}:=(\mu_{1},\mu_{0})$ and known variances
$\sigma_{1}^{2},\sigma_{0}^{2}$. It is without loss of generality
to take $\sigma_{1}^{2},\sigma_{0}^{2}$ to be known in the current
setting, as they could otherwise be completely determined in an instant
from the quadratic variations of the signal processes $x_{1}(\cdot)$
and $x_{0}(\cdot)$, defined below in (\ref{eq:diffusion=000020process=000020-=0000201}).
The aim of the decision maker (DM) is to determine which treatment
to implement on the population. To guide her choice, the DM conducts
a sequential experiment, while paying a flow cost $c$ as long as
the experiment is in progress. At each time-point $t$, the DM samples
a treatment according to the sampling rule $\pi_{a}(t)\equiv\pi(A=a\vert\mathcal{F}_{t}),a\in\{0,1\}$,
which specifies the probability of selecting treatment $a$ given
some filtration $\mathcal{F}_{t}$. The DM then keeps track of the
signals, $x_{1}(t),x_{0}(t)$ from the two treatments, as well as
the fraction of times, $q_{1}(t),q_{0}(t)$ each treatment was sampled
so far:
\begin{align}
dx_{a}(t) & =\mu_{a}\pi_{a}(t)dt+\sigma_{a}\sqrt{\pi_{a}(t)}dW_{a}(t),\label{eq:diffusion=000020process=000020-=0000201}\\
dq_{a}(t) & =\pi_{a}(t)dt.\label{eq:diffusion=000020process=000020-=0000202}
\end{align}
Here, $W_{1}(t),W_{0}(t)$ are independent one-dimensional Wiener
processes. The experiment ends in accordance with an $\mathcal{F}_{t}$-adapted
stopping time, $\tau$. At the conclusion of the experiment, the DM
chooses an $\mathcal{F}_{\tau}$ measurable implementation rule, $\delta\in\{0,1\}$,
specifying which treatment to implement on the population. The DM's
decision thus consists of the triple $\bm{d}:=(\pi,\tau,\delta)$. 

Denote $s(t)=(x_{1}(t),x_{0}(t),q_{1}(t),q_{0}(t))$ and take $\mathcal{F}_{t}\equiv\sigma\{s(u);u\le t\}$
to be the filtration generated by the state variables $s(\cdot)$
until time $t$.\footnote{As in \citet{liang2022dynamically}, we restrict attention to sampling
rules $\pi_{a}$ for which a weak solution to the functional SDEs
(\ref{eq:diffusion=000020process=000020-=0000201}), (\ref{eq:diffusion=000020process=000020-=0000202})
exists. This is true if either $\pi_{a}:\left\{ s(z):z\le t\right\} \to[0,1]$
is continuous, see \citet[Section 5.4]{karatzas2012brownian}, or,
if it is any deterministic function of $t$.} Let $\mathbb{E}_{\bm{d}\vert\bm{\mu}}[\cdot]$ denote the expectation
under a decision rule $\bm{d}$, given some value of $\bm{\mu}$.
We evaluate various decision rules by the maximum regret criterion,
defined as
\begin{align}
V_{\max}(\bm{d}) & =\max_{\bm{\mu}\in\mathbb{R}\times\mathbb{R}}V\left(\bm{d},\bm{\mu}\right),\ \textrm{with}\nonumber \\
V\left(\bm{d},\bm{\mu}\right) & :=\mathbb{E}_{\bm{d}\vert\bm{\mu}}\left[\max\{\mu_{1}-\mu_{0},0\}-(\mu_{1}-\mu_{0})\delta+c\tau\right].\label{eq:regret=000020definition}
\end{align}
To understand this expression, consider an oracle decision rule $\{\tau=0,\delta=\mathbb{I}\{\mu_{1}>\mu_{0}\}\}$,
which has full knowledge of $\bm{\mu}$. The oracle would achieve
a realized welfare of $\max\{\mu_{1},\mu_{0}\}$. In contrast, a given
decision rule $\bm{d}$ generates a realized welfare of $\mu_{0}+(\mu_{1}-\mu_{0})\delta-c\tau$.
The difference between these two welfares, $\max\{\mu_{1}-\mu_{0},0\}-(\mu_{1}-\mu_{0})\delta+c\tau$,
is referred to as regret. The quantity $V(\bm{d},\bm{\mu})$ therefore
represents the `frequentist regret', i.e., the expected regret of
$\bm{d}$ given $\bm{\mu}$. The decision rule, $\bm{d}^{*}$, that
minimizes $V_{\max}(\bm{d})$ is known as the minimax-regret optimal
decision rule. 

Minimax regret is a commonly used decision theoretic criterion when
the DM faces ambiguity over the values of $\bm{\mu}$. In contrast,
a Bayesian DM would place some prior $p_{0}$ over $\bm{\mu}$ and
aim to minimize Bayes regret, defined as 
\begin{equation}
V(\bm{d},p_{0}):=\int V\left(\bm{d},\bm{\mu}\right)dp_{0}(\bm{\mu}).\label{eq:Bayes=000020regret}
\end{equation}
We can relate max-regret to Bayes regret as $V_{\max}(\bm{d})=\sup_{p_{0}\in\mathcal{P}}V(\bm{d},p_{0})$,
where $\mathcal{P}$ denotes the set of all possible probability distributions
over $\bm{\mu}$. This suggests a multiple prior interpretation for
the minimax regret criterion. As we show in Section \ref{sec:Minimax-regret},
minimax regret can be viewed as the value of a zero-sum game played
between nature and the DM, where nature chooses the prior $p_{0}$
and the DM chooses the decision rule $\bm{d}$. The minimax-regret
optimal rule, $\bm{d}^{*}$, is then Bayes optimal under nature's
regret-maximizing choice of the prior, also known as the least-favorable
prior. 

The decision rules $\bm{d}$ are dynamic since they are history dependent.
But as stated, the max-regret criterion $V_{\max}(\bm{d})$ is `static'
since it ranks decision rules only at $t=0$; it implicitly assumes
the DM can fully commit to the course of action prescribed by $\bm{d}^ {}$.
Nonetheless, the criterion admits a dynamically consistent extension,
$V_{\max}(\bm{d};t)$, which allows for a consistent conditional ranking
of $\bm{d}$ given any history $\mathcal{F}_{t}$. This extension
is possible because the space of priors $\mathcal{P}$ is unrestricted
and therefore `rectangular' in the sense of \citet{epstein2003recursive}.
As in \citet{epstein2003recursive}, rectangularity implies existence
of a $V_{\max}(\bm{d};t)$ with a recursive structure, such that $V_{\max}(\bm{d};0)=V_{\max}(\bm{d})$,
\begin{align*}
V_{\max}(\bm{d};\tau) & =\sup_{p_{0}\in\mathcal{P}}\mathbb{E}_{p_{0}}\left[\left.\max\{\mu_{1}-\mu_{0},0\}-(\mu_{1}-\mu_{0})\delta\right|\mathcal{F}_{\tau}\right],\ \textrm{and }\\
V_{\max}(\bm{d};t) & =\sup_{p_{0}\in\mathcal{P}}\mathbb{E}_{p_{0}}\left[\left.c\cdot(\tau\wedge t^{\prime}-t)+V_{\max}\left(\bm{d};t^{\prime}\right)\right|\mathcal{F}_{t}\right]\ \forall\ t^{\prime}>t,
\end{align*}
where $\mathbb{E}_{p_{0}}\left[\left.\cdot\right|\mathcal{F}_{t}\right]$
is the expectation with respect to the posterior of $p_{0}$ given
$\mathcal{F}_{t}$. Thus, $\bm{d}^{*}$ is dynamically optimal under
$V_{\max}(\bm{d};t)$, the dynamically consistent extension of $V_{\max}(\bm{d})$.

In any event, dynamic consistency is arguably less relevant in the
context of the A/B testing examples that are the focus of this paper.
In these examples, it is quite reasonable to suppose that the DM is
able to commit to the chosen decision rule. For instance, in clinical
trials, regulatory agencies explicitly require and enforce adherence
to a pre-specified experimental strategy, see, e.g., FDA's guidance
for adaptive experiments \citep[Section III.C]{us2019guidance}. 

\subsection{Best arm identification}

The best arm identification problem is a special case of the generalized
Wald problem where the stopping time is fixed beforehand and set to
$\tau=1$ without loss of generality. This is equivalent to fixing
the number of observations before the start of the experiment; in
fact, we show in Section \ref{sec:Formal-properties} that a unit
time interval corresponds to a pre-specified number of observations,
$n$, in a discrete time analysis. Thus, decisions now consist only
of $\bm{d}=(\pi,\delta)$, but $\pi$ is still allowed to be adaptive.
If we further restrict $\pi$ to be fixed (i.e., non-adaptive), we
get back to the typical setting of Randomized Control Trials (RCTs). 

Despite these differences, we show in Section \ref{sec:Minimax-regret}
that the minimax-regret optimal sampling and implementation rules
are the same in all cases; the optimal sampling rule is the Neyman
allocation $\pi_{a}^{*}(t)=\sigma_{a}/(\sigma_{1}+\sigma_{0})$, while
the optimal implementation rule is to choose the treatment with the
higher average outcomes. Somewhat surprisingly, then, there is no
difference in the optimal strategy between best arm identification
and standard RCTs (under minimax regret). The presence of $\tau$,
however, makes the generalized Wald problem fundamentally different
from the other two. We provide a relative comparison of the benefit
of optimal stopping in Section \ref{subsec:Benefit-of-adaptive}.

\subsection{Bayesian formulation\protect\label{subsec:Bayesian-formulation}}

It is convenient to first describe minimal regret under the Bayesian
approach, given a prior $p_{0}$. As noted earlier, we can characterize
minimax regret as Bayes regret under a least-favorable prior. 

Let $p(\bm{\mu}\vert s)$ denote the posterior density of $\bm{\mu}$
given the current state $s=(x_{1},x_{0},q_{1},q_{0})\in\mathbb{R}^{4}$.
By standard results in stochastic filtering, (here, and in what follows,
$\propto$ denotes equality up to a normalization constant)
\begin{align*}
p(\bm{\mu}\vert s) & \propto p(s\vert\bm{\mu})\cdot p_{0}(\bm{\mu})\\
 & \propto p_{q_{1}}(x_{1}\vert\mu_{1})\cdot p_{q_{0}}(x_{0}\vert\mu_{0})\cdot p_{0}(\bm{\mu});\quad p_{q_{a}}(\cdot\vert\mu_{a}):=\mathcal{N}(\cdot\vert q_{a}\mu_{a},q_{a}\sigma_{a}^{2})
\end{align*}
where $\mathcal{N}(\cdot\vert\mu,\sigma^{2})$ is the normal density
with mean $\mu$ and variance $\sigma^{2}$, and the second proportionality
follows from the fact $W_{1}(\cdot),W_{0}(\cdot)$ are independent
Wiener processes. 

Define $V^{*}(s;p_{0})$ as the minimal expected Bayes regret given
state $s$, i.e.,
\[
V^{*}(s;p_{0})=\inf_{\bm{d}\in\mathcal{D}}\mathbb{E}_{\bm{\mu}\vert s}\left[V\left(\bm{d},\bm{\mu}\right)\right],
\]
where $\mathcal{D}$ is the set of all decision rules that satisfy
the measurability conditions set out previously. The minimal (ex-ante)
Bayes regret, following (\ref{eq:Bayes=000020regret}), is then related
to $V^{*}(\cdot;p_{0})$ as $\inf_{\bm{d}\in\mathcal{D}}V(\bm{d},p_{0})=V^{*}(s_{0};p_{0})$,
where $s_{0}:=(0,0,0,0)$ represents the initial state. In principle,
one could characterize $V^{*}(\cdot;p_{0})$ as a Hamilton-Jacobi-Bellman
Variational Inequality (HJB-VI; \citealp[Chapter 10]{oksendal2003stochastic}),
compute it numerically and characterize the optimal Bayes decision
rules. However, this can be computationally expensive, and moreover,
does not provide a closed form characterization of the optimal decisions.
Analytical expressions can be obtained under two types of priors:

\subsubsection{Gaussian priors}

In this case, the posterior is also Gaussian and its mean and variance
can be computed analytically. \citet{liang2022dynamically} derive
the optimal decision rule in this setting. See Section \ref{subsec:Discussion}
for a comparison with our proposal. Additional details are provided
in Appendix H. 

\subsubsection{Two-point priors}

Two point priors are closely related to hypothesis testing and the
sequential likelihood ratio procedures of \citet{wald1947sequential}
and \citet{arrow1949bayes}. More importantly for us, the least-favorable
prior for minimax regret, described in the next section, has a two
point support. 

Suppose the prior over $\bm{\mu}\equiv(\mu_{1},\mu_{0})$ is supported
on the two points $(\bar{a},\bar{b}),(\underline{a},\underline{b})$.
Let $\lambda=1$ represent the event $\bm{\mu}=(\bar{a},\bar{b})$
and $\lambda=0$ the event $\bm{\mu}=(\underline{a},\underline{b})$.
Also, let $(\Omega,\mathcal{F},\mathbb{P}_{\pi})$ denote the relevant
probability space given a (possibly) randomized policy $\pi$, where
$\mathcal{F}:=\cup_{t=1}^{\infty}\mathcal{F}_{t}\cup\sigma(\lambda)$
is the $\sigma$-field generated by $\lambda$ and the filtration
$\{\mathcal{F}_{t}\}_{t}$ defined previously, and $\mathbb{P}_{\pi}$
is the joint probability distribution over $\bm{\mu}$ and the sample
paths of $s(t)$ under $\pi$. Set $P_{\pi}^{0},P_{\pi}^{1}$ to be
the probability measures $P_{\pi}^{0}(A):=\mathbb{P}_{\pi}(A\vert\lambda=0)$
and $P_{\pi}^{1}(A):=\mathbb{P}_{\pi}(A\vert\lambda=1)$ for any $A\in\mathcal{F}$. 

Clearly, the likelihood ratio process $\varphi^{\pi}(t):=\mathbb{E}_{P_{\pi}^{0}}\left[\left.\frac{dP_{\pi}^{1}}{dP_{\pi}^{0}}\right|\mathcal{F}_{t}\right]$
is a sufficient statistic for $\lambda$.\footnote{Note that $\frac{dP_{\pi}^{1}}{dP_{\pi}^{0}}$ is a random variable,
being the Radon-Nikodym derivative of $P_{\pi}^{1}$ with respect
to $P_{\pi}^{0}$. } An application of the Girsanov theorem, noting that $W_{1}(\cdot),W_{0}(\cdot)$
are independent of each other, gives (see also \citealp[Section 4.2.1]{shiryaev2007optimal})
\begin{align}
\ln\varphi^{\pi}(t) & =\frac{(\bar{a}-\underline{a})}{\sigma_{1}^{2}}x_{1}(t)+\frac{(\bar{b}-\underline{b})}{\sigma_{0}^{2}}x_{0}(t)-\frac{(\bar{a}^{2}-\underline{a}^{2})}{2\sigma_{1}^{2}}q_{1}(t)-\frac{(\bar{b}^{2}-\underline{b}^{2})}{2\sigma_{0}^{2}}q_{0}(t).\label{eq:LR=000020process}
\end{align}
Let $m_{0}$ denote the prior probability that $\lambda=1$. Additionally,
given a sampling rule $\pi$, let $m^{\pi}(t)=\mathbb{P}(\lambda=1\vert\mathcal{F}_{t})$
denote the belief process describing the posterior probability that
$\lambda=1$. Following \citet[Section 4.2.1]{shiryaev2007optimal},
$m^{\pi}(t)$ can be related to $\varphi^{\pi}(t)$ as 
\begin{equation}
m^{\pi}(t)=\frac{m_{0}\varphi^{\pi}(t)}{(1-m_{0})+m_{0}\varphi^{\pi}(t)}.\label{eq:belief=000020process}
\end{equation}

The Bayes optimal implementation rule at the end of the experiment
is
\begin{align}
\delta^{\pi,\tau} & =\mathbb{I}\left\{ \bar{a}m^{\pi}(\tau)+\underline{a}(1-m^{\pi}(\tau))\ge\bar{b}m^{\pi}(\tau)+\underline{b}(1-m^{\pi}(\tau))\right\} \nonumber \\
 & =\mathbb{I}\left\{ \ln\varphi^{\pi}(\tau)\ge\ln\frac{(\underline{b}-\underline{a})(1-m_{0})}{(\bar{a}-\bar{b})m_{0}}\right\} .\label{eq:allocation=000020rule}
\end{align}
The superscript on $\delta$ highlights that the above implementation
rule is conditional on a given choice of $(\pi,\tau)$. Relatedly,
the Bayes regret at the implementation phase of the experiment (from
employing the optimal implementation rule) is
\begin{align}
\varpi^{\pi}(\tau) & :=\mathbb{E}_{\mathbb{P}_{\pi}}\left[\left.\max\{\mu_{1}-\mu_{0},0\}-(\mu_{1}-\mu_{0})\delta^{\pi,\tau}\right|\mathcal{F}_{\tau}\right]\nonumber \\
 & =\mathbb{E}_{\mathbb{P}_{\pi}}\left[\left.\max\{\mu_{1}-\mu_{0},0\}\right|\mathcal{F}_{\tau}\right]-\mathbb{E}_{\mathbb{P}_{\pi}}\left[\left.\mu_{1}-\mu_{0}\right|\mathcal{F}_{\tau}\right]\delta^{\pi,\tau}\nonumber \\
 & =\min\left\{ (\bar{a}-\bar{b})m^{\pi}(\tau),(\underline{b}-\underline{a})(1-m^{\pi}(\tau))\right\} .\label{eq:Bayes=000020regret=000020at=000020end=000020of=000020experiment}
\end{align}
Hence, for a given sampling rule $\pi$, the Bayes optimal stopping
time $\tau^{\pi}$, can be obtained as the solution to the optimal
stopping problem
\begin{equation}
\tau^{\pi}=\inf_{\tau\in\mathcal{T}}\mathbb{E}_{\pi}\left[\varpi^{\pi}(\tau)+c\tau\right],\label{eq:Bayes=000020optimal=000020stopping=000020rule}
\end{equation}
where $\mathcal{T}$ is the set of all $\mathcal{F}_{t}$ measurable
stopping times, and $\mathbb{E}_{\pi}[\cdot]$ denotes the expectation
under the sampling rule $\pi$. 

\section{Minimax regret and optimal decision rules\protect\label{sec:Minimax-regret}}

The minimax regret value can be written as
\begin{equation}
\inf_{\bm{d}\in\mathcal{D}}V_{\max}(\bm{d})=\inf_{\bm{d}\in\mathcal{D}}\sup_{p_{0}\in\mathcal{P}}V(\bm{d},p_{0}).\label{eq:def=000020of=000020minimax=000020risk}
\end{equation}
Following \citet{wald1945statistical}, we can characterize minimax
regret as the value of a zero-sum game played between nature and the
DM. Nature's action involves choosing a prior $p_{0}\in\mathcal{P}$
over $\bm{\mu}$, while the DM chooses the decision rule $\bm{d}$.
The equilibrium action of nature is termed the least-favorable prior,
and that of the DM, the minimax decision rule. Note that nature's
action in the game is static, as it only chooses a prior $p_{0}$
at the beginning of the experiment. In contrast, the DM selects a
dynamic decision rule $\bm{d}$ that, to be a best response to nature's
choice, must be Bayes optimal with respect to that choice of prior.
Consequently, the minimax-regret optimal rule $\bm{d}^{*}$ must be
consistent with Bayesian updating of the least-favorable prior throughout
the experiment. 

The following is the main result of this section: Let $\gamma_{0}^{*}\approx0.536357$,
$\Delta_{0}^{*}\approx2.19613$ denote universal constants derived
from solving a univariate minimax problem (\ref{eq:minimax=000020problem=000020-=000020definition})
described later in this section. Also, define $\eta:=\left(\frac{2c}{\sigma_{1}+\sigma_{0}}\right)^{1/3}$,
$\gamma^{*}=\gamma_{0}^{*}/\eta$ and $\Delta^{*}=\eta\Delta_{0}^{*}$. 

\begin{thm}\label{theorem=0000201} The zero-sum two player game
(\ref{eq:def=000020of=000020minimax=000020risk}) has a Nash equilibrium
with a unique minimax-regret value. The minimax-regret optimal decision
rule is $\bm{d}^{*}:=(\pi^{*},\tau^{*},\delta^{*})$, where $\pi_{a}^{*}(t)=\sigma_{a}/(\sigma_{1}+\sigma_{0})$
for $a\in\{0,1\}$, 
\[
\tau^{*}=\inf\left\{ t:\left|\frac{x_{1}(t)}{\sigma_{1}}-\frac{x_{0}(t)}{\sigma_{0}}\right|\ge\gamma^{*}\right\} ,
\]
and $\delta^{*}=\mathbb{I}\left\{ \frac{x_{1}(\tau^{*})}{\sigma_{1}}-\frac{x_{0}(\tau^{*})}{\sigma_{0}}\ge0\right\} $.
Furthermore, the least-favorable prior is a symmetric two-point distribution
supported on $(\sigma_{1}\Delta^{*}/2,-\sigma_{0}\Delta^{*}/2),(-\sigma_{1}\Delta^{*}/2,\sigma_{0}\Delta^{*}/2)$.
\end{thm}

Theorem \ref{theorem=0000201} makes no claim as to the uniqueness
of the Nash equilibrium.\footnote{In fact, this would depend on the topology defined over $\mathcal{D}$
and $\mathcal{P}$.} Even if multiple equilibria were to exist, however, the value of
the game $V^{*}=\inf_{\bm{d}\in\mathcal{D}}\sup_{p_{0}\in\mathcal{P}}V(\bm{d},p_{0})$
would be unique, and $\bm{d}^{*}$ would still be minimax-regret optimal.

The optimal strategies under best arm identification can be derived
in the same manner as Theorem \ref{theorem=0000201}, but the proof
is simpler as it does not involve a stopping rule. Let $\Phi(\cdot)$
denote the CDF of the standard normal distribution. 

\begin{cor}\label{Corollary=0000201} The minimax-regret optimal
decision rule for best arm identification is $\bm{d}_{\textrm{BAI}}^{*}:=(\pi^{*},\delta^{*})$,
where $\pi^{*},\delta^{*}$ are defined in Theorem \ref{theorem=0000201}.
The corresponding least-favorable prior is a symmetric two-point distribution
supported on $(\sigma_{1}\bar{\Delta}_{0}^{*}/2,-\sigma_{0}\bar{\Delta}_{0}^{*}/2),(-\sigma_{1}\bar{\Delta}_{0}^{*}/2,\sigma_{0}\bar{\Delta}_{0}^{*}/2)$,
where $\bar{\Delta}_{0}^{*}:=2\argmax_{\delta}\delta\Phi(-\delta)$.
\end{cor}

\subsection{Proof sketch of Theorem \ref{theorem=0000201}\protect\label{subsec:Intuition-behind-Theorem}}

The main challenge with analyzing the game (\ref{eq:def=000020of=000020minimax=000020risk})
is that the action spaces $\mathcal{P},\mathcal{D}$ of both nature
and the DM are infinite dimensional. Therefore, to prove Theorem \ref{theorem=0000201},
we first restrict the action spaces of both players and then show
that a Nash equilibrium exists within this restricted class.

For nature, we employ the restricted action space, $\mathcal{P}_{\textrm{rest}}:=\{p_{\Delta}:\Delta\in\mathbb{R}^{+}\}$,
consisting of all `indifference priors' indexed by $\Delta\in\mathbb{R}$.
Specifically, each `indifference prior' $p_{\Delta}$ is a two-point
prior supported on $(\sigma_{1}\Delta/2,-\sigma_{0}\Delta/2),(-\sigma_{1}\Delta/2,\sigma_{0}\Delta/2$),
with a prior probability of $0.5$ at each support point. As for the
DM, we employ the restricted action space 
\[
\mathcal{D}_{\textrm{rest}}:=\left\{ \tilde{\bm{d}}_{\gamma}=(\pi^{*},\tau_{\gamma},\delta^{\tau_{\gamma}}):\gamma\in\mathbb{R}^{+}\right\} ,
\]
where
\begin{align}
\tau_{\gamma} & :=\inf\left\{ t:\left|\frac{x_{1}(t)}{\sigma_{1}}-\frac{x_{0}(t)}{\sigma_{0}}\right|\ge\gamma\right\} ,\ \textrm{and}\label{eq:stopping=000020time:=000020d_gamma}\\
\delta^{\tau_{\gamma}} & :=\mathbb{I}\left\{ \frac{x_{1}(\tau_{\gamma})}{\sigma_{1}}-\frac{x_{0}(\tau_{\gamma})}{\sigma_{0}}\ge0\right\} .\label{eq:implementation=000020rule:=000020d_gamma}
\end{align}

We demonstrate that for each $p_{\Delta}\in\mathcal{P}_{\textrm{rest}}$,
there exists a unique $\gamma\in\mathbb{R}^{+}$ such that $\tilde{\bm{d}}_{\gamma}$
is an unconstrained best response of the DM to $p_{\Delta}$. In other
words, $\tilde{\bm{d}}_{\gamma}$ is a best response within the unrestricted
class $\mathcal{D}$, not just within $\mathcal{D}_{\textrm{rest}}$.
Similarly, for each $\tilde{\bm{d}}_{\gamma}\in\mathcal{D}_{\textrm{rest}}$,
there exists a $\Delta\in\mathbb{R}^{+}$ such that $p_{\Delta}$
is an unconstrained best response of nature to $\tilde{\bm{d}}_{\gamma}$.
These results imply that any Nash equilibrium within the restricted
action space $\mathcal{P}_{\textrm{rest}}\times\mathcal{D}_{\textrm{rest}}$
would also be a Nash equilibrium within the unrestricted action space
$\mathcal{P}\times\mathcal{D}$. We then formally demonstrate the
existence of a Nash equilibrium in the restricted setting and characterize
the equilibrium set of actions. 

We elaborate on these steps below:

\subsubsection*{The DM's response to $p_{\Delta}$. }

The term `indifference priors' indicates that these priors make the
DM indifferent between any sampling rule $\pi$. The intuitive explanation
is as follows: Let $\lambda=1$ represent the event $\bm{\mu}=(\sigma_{1}\Delta/2,-\sigma_{0}\Delta/2)$
and $\lambda=0$ the event $\bm{\mu}=(-\sigma_{1}\Delta/2,\sigma_{0}\Delta/2)$.
These support points are configured in such a way that both treatments
provide equal information about $\lambda$, making the choice of treatment
irrelevant. To illustrate, assume $\sigma_{1}=\sigma_{0}=1$. If the
DM samples arm 1 for a period of time $\delta t$, she would observe
the signal process $x_{1}(t)=(2\lambda-1)\frac{\Delta}{2}t+W_{1}(t)$
over that time-period, with a drift of either $\Delta/2$ or $-\Delta/2$
depending on whether $\lambda=1$ or $\lambda=0$. Alternatively,
sampling arm 0 yields $x_{0}(t)=-(2\lambda-1)\frac{\Delta}{2}t+W_{0}(t)$,
with an exactly opposite drift. Since $W_{1}(\cdot),W_{0}(\cdot)$
are independent Wiener processes, both sampling strategies are equally
informative about $\lambda$ in the Blackwell sense.

We now describe the formal argument. For both support points of $p_{\Delta}$,
(\ref{eq:LR=000020process}) implies 
\begin{equation}
\ln\varphi^{\pi}(t)=\left(\frac{x_{1}(t)}{\sigma_{1}}-\frac{x_{0}(t)}{\sigma_{0}}\right)\cdot\Delta.\label{eq:LR=000020process=000020indiffierence}
\end{equation}
Suppose $\lambda=1$. By (\ref{eq:diffusion=000020process=000020-=0000201})
and (\ref{eq:diffusion=000020process=000020-=0000202}) 
\begin{align}
\frac{dx_{1}(t)}{\sigma_{1}}-\frac{dx_{0}(t)}{\sigma_{0}} & =\frac{\Delta}{2}dt+\sqrt{\pi_{1}(t)}dW_{1}(t)-\sqrt{\pi_{0}(t)}dW_{0}(t)\nonumber \\
 & =\frac{\Delta}{2}dt+d\tilde{W}(t),\label{eq:LR=000020evolution}
\end{align}
where $\tilde{W}(\cdot)$, defined as $d\tilde{W}(t):=\sqrt{\pi_{1}(t)}dW_{1}(t)-\sqrt{\pi_{0}(t)}dW_{0}(t)$,
is a one dimensional Wiener process, being a linear combination of
two independent Wiener processes with $\pi_{1}(t)+\pi_{0}(t)=1$.
Plugging the above into (\ref{eq:LR=000020process=000020indiffierence})
gives 
\[
d\ln\varphi^{\pi}(t)=\frac{\Delta^{2}}{2}dt+\Delta d\tilde{W}(t).
\]
In a similar manner, we can show under $\lambda=0$ that $d\ln\varphi^{\pi}(t)=-\frac{\Delta^{2}}{2}dt+\Delta d\tilde{W}(t).$
Thus, the evolution of the log-likelihood ratio process, $\ln\varphi^{\pi}(t)$,
can be decomposed into two parts: a drift term $(2\lambda-1)\frac{\Delta^{2}}{2}dt$
that depends on the state of the world $\lambda\in\{0,1\}$, and noise
$\Delta d\tilde{W}(t)$. Different sampling rules, $\pi$, induce
the same drift and leave unchanged the distribution of noise, $\tilde{W}(\cdot)$.
Therefore, the choice of $\pi$ does not affect the sample-path distribution
of $\varphi^{\pi}(\cdot)$, and consequently, has no bearing on the
sample-path distribution of the belief process $m^{\pi}(\cdot)$.

Crucially, this invariance to the choice of $\pi$ holds at every
time point during the experiment, even as $p_{\Delta}$ is revised
through Bayesian updating. The key to this invariance lies in the
fact that Bayesian updating does not alter the support points of the
prior, and it is solely these support points, together with $\pi$,
that govern the evolution of $\varphi^{\pi}(\cdot)$ under Wiener
process noise. Now, the precise form of the support points of $p_{\Delta}$
ensures the drift of $\varphi^{\pi}(t)$ is independent of $\pi$.
Then, the linearity property of Wiener processes implies that a linear
combination of such processes remains a Wiener process, thereby preserving
the independence of the noise process from the choice of $\pi$ as
well. 

As the distributions of $\varphi^{\pi}(\cdot),m^{\pi}(\cdot)$ do
not depend on $\pi$, the Bayes optimal stopping time in (\ref{eq:Bayes=000020optimal=000020stopping=000020rule})
is also independent of $\pi$ for indifference priors (standard results
in optimal stopping, see e.g., \citealp[Chapter 10]{oksendal2003stochastic},
imply that the optimal stopping time in (\ref{eq:Bayes=000020optimal=000020stopping=000020rule})
is a function only of $m^{\pi}(t)$ which is now independent of $\pi$).
In fact, it has the same form as the optimal stopping time in the
Bayesian hypothesis testing problem of \citet{arrow1949bayes}, analyzed
in continuous time by \citet[Section 4.2.1]{shiryaev2007optimal}
and \citet{morris2019wald}. An adaptation of their results (see,
Lemma \ref{Lemma=0000201} in Appendix \ref{sec:Appendix:A}) shows
that the Bayes optimal stopping time corresponding to $p_{\Delta}$
is
\begin{equation}
\tau_{\gamma(\Delta)}=\inf\left\{ t:\left|\frac{x_{1}(t)}{\sigma_{1}}-\frac{x_{0}(t)}{\sigma_{0}}\right|\ge\gamma(\Delta)\right\} ,\label{eq:definition=000020of=000020tau_delta}
\end{equation}
where $\gamma(\Delta)$ is defined in Lemma \ref{Lemma=0000201}.
By (\ref{eq:allocation=000020rule}) and (\ref{eq:LR=000020process=000020indiffierence}),
the corresponding Bayes optimal implementation rule is seen to be
$\delta^{\tau_{\gamma(\Delta)}}$, as defined in (\ref{eq:implementation=000020rule:=000020d_gamma}).
Hence, the decision rule $\tilde{\bm{d}}_{\gamma(\Delta)}$ is a best
response of the DM to nature's choice of $p_{\Delta}$. 

\subsubsection*{Nature's response to $\tilde{\bm{d}}_{\gamma}$. }

Lemma \ref{Lemma=0000202} in Appendix \ref{sec:Appendix:A} shows
that the frequentist regret $V\left(\tilde{\bm{d}}_{\gamma},\bm{\mu}\right)$,
given some $\bm{\mu}=(\mu_{1},\mu_{0})$, depends only on $\vert\mu_{1}-\mu_{0}\vert$.
To understand this result, observe that 
\begin{equation}
V\left(\tilde{\bm{d}}_{\gamma},\bm{\mu}\right)=\max\{\mu_{1}-\mu_{0},0\}-(\mu_{1}-\mu_{0})\mathbb{E}_{\bm{d}\vert\bm{\mu}}\left[\delta^{\tau_{\gamma}}\right]+c\mathbb{E}_{\bm{d}\vert\bm{\mu}}\left[\tau_{\gamma}\right].\label{eq:frequentist=000020regret=000020expansion}
\end{equation}
Clearly, $\tau_{\gamma},\delta^{\tau_{\gamma}}$ depend on the data
only through the stochastic process $\sigma_{1}^{-1}x_{1}(\cdot)-\sigma_{0}^{-1}x_{0}(\cdot)$.
Under the Neyman allocation, (\ref{eq:diffusion=000020process=000020-=0000201})
and (\ref{eq:diffusion=000020process=000020-=0000202}) imply 
\begin{equation}
\frac{x_{1}(t)}{\sigma_{1}}-\frac{x_{0}(t)}{\sigma_{0}}=\frac{\mu_{1}-\mu_{0}}{\sigma_{1}+\sigma_{0}}t+\tilde{W}(t)\label{eq:evolution=000020of=000020rho}
\end{equation}
for any $\bm{\mu}\in\mathbb{R}^{2}$, where $\tilde{W}(\cdot):=\sqrt{\frac{\sigma_{1}}{\sigma_{1}+\sigma_{0}}}W_{1}(\cdot)-\sqrt{\frac{\sigma_{0}}{\sigma_{1}+\sigma_{0}}}W_{0}(\cdot)$
is a standard one-dimensional Wiener process. Consequently, the distributions
of $\tau_{\gamma},\delta^{\tau_{\gamma}}$ depend on $\bm{\mu}$ only
through $\mu_{1}-\mu_{0}$. This in turn implies, based on (\ref{eq:frequentist=000020regret=000020expansion}),
that $V\left(\tilde{\bm{d}}_{\gamma},\bm{\mu}\right)$ depends on
$\bm{\mu}$ only through $\mu_{1}-\mu_{0}$. This dependence can be
further reduced to $\vert\mu_{1}-\mu_{0}\vert$ by symmetry since
interchanging the treatment labels $0,1$ would not affect the frequentist
regret of $\tilde{\bm{d}}_{\gamma}$. 

Since $\bm{\mu}$ affects $V\left(\tilde{\bm{d}}_{\gamma},\bm{\mu}\right)$
only through $\vert\mu_{1}-\mu_{0}\vert$, it is maximized at $\vert\mu_{1}-\mu_{2}\vert=(\sigma_{1}+\sigma_{0})\Delta(\gamma)/2$,
where $\Delta(\gamma)$ is some function of $\gamma$. Thus, the best
response of nature to $\tilde{\bm{d}}_{\gamma}$ is to pick any prior
that is supported on $\left\{ \bm{\mu}:\vert\mu_{1}-\mu_{0}\vert=(\sigma_{1}+\sigma_{0})\Delta(\gamma)/2\right\} $.
Therefore, the two-point prior $p_{\Delta(\gamma)}$ is a best response
to $\tilde{\bm{d}}_{\gamma}$. 

The use of Neyman allocation is essential for the above conclusion.
With a different sampling rule, the distributions of $\tau_{\gamma},\delta^{\tau_{\gamma}}$
would depend not only on $\mu_{1}-\mu_{0}$, but also on the individual
levels of $\mu_{1},\mu_{0}$. In such cases, nature could drive the
max-regret of the corresponding decision rule to $\infty$ through
an adversarial choice of $\bm{\mu}$, as we show in Appendix B. This
explains why only the Neyman allocation is minimax optimal, even though
the DM is indifferent to any sampling rule under $p_{\Delta}$: it
is needed to ensure nature's choice of $p_{\Delta}$ is supported
as a best response to $\tilde{\bm{d}}_{\gamma}$. 

\subsubsection*{Nash equilibrium. }

The above observations imply that the overall Nash equilibrium to
(\ref{eq:def=000020of=000020minimax=000020risk}) is the same as the
Nash equilibrium in the restricted sub-problem where nature chooses
an indifference prior, $p_{\Delta}$, indexed by $\Delta\in\mathbb{R}^{+}$,
and the DM chooses a decision rule $\tilde{\bm{d}}_{\gamma}$, indexed
by $\gamma\in\mathbb{R}^{+}$. Thus, the action spaces of nature and
the DM in this sub-problem are scalar. Lemma \ref{Lemma=0000203}
in Appendix \ref{sec:Appendix:A} formally demonstrates existence
of a Nash equilibrium in the sub-problem using Sion's minimax theorem
\citep{sion1958general}. The equilibrium values of $\Delta,\gamma$
can be computed numerically by writing down the relevant first order
conditions for a Nash equilibrium (see, also, Figure \ref{fig:pf=000020Thm1}
for the best response functions). The universal constants, $\gamma_{0}^{*},\Delta_{0}^{*}$
used in Theorem \ref{theorem=0000201} are derived in this manner.
Specifically, Lemma \ref{Lemma=0000203} demonstrates that these constants
solve the following minimax problem, which characterizes the Nash
equilibrium in the restricted sub-problem when $\eta=1$: 
\begin{equation}
\min_{\gamma\in\mathbb{R}^{+}}\max_{\Delta\in\mathbb{R}^{+}}\left\{ \Delta\frac{1-e^{-\Delta\gamma}}{e^{\Delta\gamma}-e^{-\Delta\gamma}}+\frac{2\gamma}{\Delta}\frac{e^{\Delta\gamma}+e^{-\Delta\gamma}-2}{e^{\Delta\gamma}-e^{-\Delta\gamma}}\right\} .\label{eq:minimax=000020problem=000020-=000020definition}
\end{equation}

\subsection{Discussion\protect\label{subsec:Discussion}}

\subsubsection{Sampling rule}

Perhaps the most striking aspect of the sampling rule is that it is
just the Neyman allocation. This rule is non-adaptive (i.e., it is
history independent), and is also independent of sampling costs. In
fact, Corollary \ref{Corollary=0000201} shows that the sampling and
implementation rules are identical to those used in the best arm identification
problem.

The Neyman allocation is well known in the RCT literature for being
the sampling rule that minimizes the estimation variance of the treatment
effect $\mu_{1}-\mu_{0}$. \citet{armstrong2022asymptotic} shows
that it retains its optimality for estimating $\mu_{1}-\mu_{0}$ even
when adaptive sampling strategies are allowed. While Armstrong's \citeyearpar{armstrong2022asymptotic}
result does not apply to best arm identification, Corollary \ref{Corollary=0000201}
confirms that Neyman allocation is optimal in this context as well.
Hence, practitioners should continue using the randomization designs
employed in standard (i.e., non-sequential) experiments, even when
adaptivity is allowed. 

By way of comparison, the optimal sampling rule under Gaussian priors
is also non-adaptive, but it varies deterministically with time \citep{liang2022dynamically}.
In fact, after a set time point $t^{*}$ that depends on $(\sigma_{1},\sigma_{0})$
and the prior variance, it too becomes equal to the Neyman allocation;
see Appendix H for a detailed description. There are likely priors
for which the Bayes optimal sampling rule is adaptive, but analyzing
optimal decision rules under general classes of priors (beyond Gaussian
or two-point priors) presents a challenging stochastic-filtering problem.
As such, it appears difficult to provide any general claims on what
priors lead to adaptive sampling.

\subsubsection{Stopping time}

The stopping time $\tau^{*}$ is adaptive but has a simple form: the
DM ends the experiment when $\rho(t):=\sigma_{1}^{-1}x_{1}(t)-\sigma_{0}^{-1}x_{0}(t)$
exceeds $(\frac{\sigma_{1}+\sigma_{0}}{2c})^{1/3}\gamma_{0}^{*}$
in absolute value. The threshold is decreasing in $c$ and increasing
in $\sigma_{1}+\sigma_{0}$. Let $\bar{x}_{a}(t):=x_{a}(t)/q_{a}(t)$
denote the sample average of outcomes from treatment $a$ at time
$t$. Since $q_{a}(t)=\sigma_{a}t/(\sigma_{1}+\sigma_{0})$ under
$\pi^{*}$, we can rewrite the optimal stopping rule as 
\[
\tau^{*}=\inf\left\{ t:t\left|\bar{x}_{1}(t)-\bar{x}_{0}(t)\right|\ge(\sigma_{1}+\sigma_{0})\gamma^{*}\right\} ,
\]
meaning the experiment is stopped when the difference in average outcomes
multiplied by the duration $t$ exceeds $(\sigma_{1}+\sigma_{0})\gamma^{*}$.
Furthermore, from the definition of $\tau^{*}$ and (\ref{eq:evolution=000020of=000020rho}),
we can infer that earlier stopping is indicative of larger reward
gaps $\mu_{1}-\mu_{0}$, with the average length of the experiment
being longest when $\mu_{1}-\mu_{0}=0$. 

In contrast, \citet{fudenberg2018speed} show that when $\sigma_{1}=\sigma_{0}$,
the Bayes optimal stopping time under the independent Gaussian prior
$\bm{\mu}\sim\mathcal{N}(0,\varsigma)\times\mathcal{N}(0,\varsigma)$
has the form $\tau_{\textrm{Bayes}}=\mathbb{I}\left\{ \vert\rho(t)\vert\ge b^{*}(t;c,\sigma_{1},\varsigma)\right\} $,
where the threshold $b^{*}(t;\cdot)$ is now time-varying. The following
intuition, adapted from \citet{fudenberg2018speed}, helps explain
the difference: Suppose that $\rho(t)\approx0$ for some large $t$.
Under the Gaussian prior, this likely indicates that $\mu_{1}-\mu_{0}$
is close to $0$, suggesting no significant difference between the
treatments, so the DM should terminate the experiment straightaway.
Conversely, under the least-favorable prior $p_{\Delta^{*}}$, which
has a two-point support, $\rho(t)\approx0$ would be interpreted as
noise, so the DM should proceed henceforth as if starting the experiment
from scratch. Thus, the properties of the stopping time are very different
depending on the prior. The above intuition also suggests that the
relation between $\mu_{1}-\mu_{0}$ and stopping times is more complicated
under Gaussian priors, and not monotone as under minimax regret. 

The stopping time, $\tau^{*}$, induces a specific probability of
mis-identification of the optimal treatment under the least-favorable
prior. By Lemmas \ref{Lemma=0000202} and \ref{Lemma=0000203}, this
probability is 
\begin{equation}
\alpha^{*}=\frac{1-e^{-\Delta^{*}\gamma^{*}}}{e^{\Delta^{*}\gamma^{*}}-e^{-\Delta^{*}\gamma^{*}}}=\frac{1-e^{-\Delta_{0}^{*}\gamma_{0}^{*}}}{e^{\Delta_{0}^{*}\gamma_{0}^{*}}-e^{-\Delta_{0}^{*}\gamma_{0}^{*}}}\approx0.235.\label{eq:mis-identification=000020error}
\end{equation}
Interestingly, $\alpha^{*}$ is independent of the model parameters
$c,\sigma_{1},\sigma_{0}$. This is because the least-favorable prior
adjusts the reward gap in response to these quantities.

Another remarkable property, following from \citet[Theorem 1]{fudenberg2018speed},
is that the probability of mis-identification is independent of the
stopping time for any given value of $\bm{\mu}$, i.e., $\mathbb{P}(\delta^{*}=1\vert\tau^{*},\bm{\mu}=\bm{b})=\mathbb{P}(\delta^{*}=1\vert\bm{\mu}=\bm{b})$
for any $\bm{b}\in\mathbb{R}^{2}$. This is again different from the
setting with Gaussian priors, where earlier stopping is indicative
of a higher probability of selecting the best treatment.

\subsection{Benefit of adaptive experimentation\protect\label{subsec:Benefit-of-adaptive}}

In both best arm identification and standard RCTs, the number of units
of experimentation is specified beforehand. As we have seen previously,
the Neyman allocation is minimax optimal under both adaptive and non-adaptive
experiments. The benefit of the decision rule, $\bm{d}^{*}$, however,
is that it enables one to stop the experiment early, thus saving on
experimental costs. To quantify this benefit, fix some values of $\sigma_{1},\sigma_{0},c$,
and suppose that nature chooses the least-favorable prior, $p_{\Delta^{*}}$,
for the generalized Wald problem. Note that $p_{\Delta^{*}}$ is in
general different from the least-favorable prior for the best arm
identification problem.\footnote{However, the two coincide if the parameter values are such that $\eta:=\left(\frac{2c}{\sigma_{1}+\sigma_{0}}\right)^{1/3}=\bar{\Delta}_{0}^{*}/\Delta_{0}^{*}\approx0.484$,
where $\Delta_{0}^{*},\bar{\Delta}_{0}^{*}$ are universal constants
defined in the contexts of Theorem \ref{theorem=0000201} and Corollary
\ref{Corollary=0000201}. } 

Let 
\[
R^{*}:=\int\mathbb{E}_{\bm{d}^{*}\vert\bm{\mu}}\left[\max\{\mu_{1}-\mu_{0},0\}-(\mu_{1}-\mu_{0})\delta\right]dp_{\Delta^{*}}
\]
denote the Bayes regret, under $p_{\Delta^{*}}$, of the minimax decision
rule $\bm{d}^{*}$ net of sampling costs. In fact, by symmetry, the
above is also the frequentist regret of $\bm{d}^{*}$ under both the
support points of $p_{\Delta^{*}}$. Now, let $T_{R^{*}}$ denote
the duration of time required in a non-adaptive experiment to achieve
the same Bayes regret $R^{*}$ (also under the least-favorable prior
and net of sampling costs). Then, making use of some results from
\citet[Section 4.2.5]{shiryaev2007optimal}, we show in Appendix B.2
that 
\begin{equation}
\frac{\mathbb{E}[\tau^{*}]}{T_{R^{*}}}=\frac{1-2\alpha^{*}}{2\left(\Phi^{-1}(1-\alpha^{*})\right)^{2}}\ln\frac{1-\alpha^{*}}{\alpha^{*}}\approx0.6.\label{eq:gain=000020from=000020adaptivity}
\end{equation}

In other words, the use of an adaptive stopping time enables us to
attain the same regret with $40\%$ fewer observations on average.
Interestingly, the above result is independent of $\sigma_{1},\sigma_{0},c$,
though the values of $\mathbb{E}[\tau^{*}]$ and $T_{R^{*}}$ do depend
on these quantities (it is only the ratio that is constant). Admittedly,
(\ref{eq:gain=000020from=000020adaptivity}) does not quantify the
welfare gain from using an adaptive experiment - this will depend
on the sampling costs - but it is nevertheless useful as an informal
measure of how much the amount of experimentation can be reduced. 

\section{Parametric regimes and small cost asymptotics\protect\label{sec:Formal-properties} }

We now turn to the analysis of parametric models in discrete time.
As before, the DM is tasked with selecting a treatment for implementation
across a population. To this end, the DM experiments sequentially
in periods $j=1,2,\dots$ after paying an `effective sampling cost'
$C$ per period. Let $1/n$ denote the time interval between successive
time periods. To analyze asymptotic behavior in this context, we introduce
small cost asymptotics, wherein $C=c/n^{3/2}$ for some $c\in(0,\infty)$,
and $n\to\infty$.

Are small cost asymptotics realistic? We contend they are, as $C$
is not the actual cost of experimentation, but rather characterizes
the tradeoff between these costs and the benefits from full-scale
implementation following the experiment. Indeed, one way to motivate
this asymptotic regime is to imagine there are $n^{3/2}$ population
units in the implementation phase, so the benefit of applying treatment
$a$ is $n^{3/2}\mu_{a}$, but we divide by $n^{3/2}$ throughout.
The actual cost of sampling an additional unit is $c$, which becomes
$c/n^{3/2}$ after the division. Moreover, time $t$ is measured in
units of $n$. This framework aligns with the practical observation
that sampling costs are relatively small compared to population size,
as seen in both online platforms (Deng et al., 2013) and clinical
trials.

For example, in Phase 3 clinical trials, the per-unit cost of treatment
is relatively high - \citet{moore2018estimated} estimate the median
cost per patient to be around 41,000\$. However, the potential welfare
implications for the population are even more substantial, as the
decisions from these trials impact millions of people and firms can
expect to earn billions of dollars from successful blockbuster drugs.
The effective marginal cost of each observation, obtained by dividing
the monetary cost by the population size, is therefore quite small
and falls well within our asymptotic framework. More generally, our
scaling suggests that if the population size is $n^{3/2}$, one should
aim to experiment on a sample size of the order $n$ to achieve optimal
welfare. This naturally leads to small cost asymptotics.

As with any asymptotic regime, small-cost asymptotics only provide
an approximation to the finite sample properties of decision rules
(unless the outcomes are truly Gaussian, in which case they would
be exact). The actual finite sample performance needs to be assessed
using simulations. Nonetheless, asymptotic analysis offers a valuable
benchmark: while there may exist decision rules that outperform our
proposal in finite samples, it would be difficult to justify using
one that is asymptotically inefficient.

\subsection{Setup in parametric regimes}

In each period, the DM assigns a treatment to a single unit of observation
according to some sampling rule $\pi_{j}(\cdot)$. The treatment assignment
is a random draw $A_{j}\sim\textrm{Bernoulli}(\pi_{j})$. This results
in an outcome $Y^{(a)}\sim P_{\theta}^{(a)}$, with $P_{\theta}^{(a)}$
denoting the population distribution of outcomes under treatment $a$.
In this section, we assume that this distribution is known up to some
unknown $\theta^{(a)}\in\mathbb{R}^{d}$. It is without loss of generality
to assume $Y^{(1)},Y^{(0)}$ are mutually independent (conditional
on $\theta^{(1)},\theta^{(0)}$) as we only ever observe the outcomes
from one treatment anyway. After observing the outcome, the DM can
decide either to stop sampling, or call up the next unit. At the end
of the experiment, the DM prescribes a treatment to apply on the population. 

We use the `stack-of-rewards-representation' for the outcomes from
each arm (\citealp[Section 4.6]{lattimore2020bandit}). Specifically,
$Y_{i}^{(a)}$ denotes the outcome for $i$-th data point corresponding
to treatment $a$. Also, ${\bf y}_{nq}:=\{Y_{i}^{(a)}\}_{i=1}^{\left\lfloor nq\right\rfloor }$
denotes the sequence of outcomes after $\left\lfloor nq\right\rfloor $
observations from treatment $a$. We can imagine that prior to the
experiment, nature draws an infinite stack of outcomes, ${\bf y}^{(a)}:=\{Y_{i}^{(a)}\}_{i=1}^{\infty}$,
corresponding to each treatment $a$, and at each period $j$, if
$A_{j}=a$, the DM observes the outcome at the top of the stack (this
outcome is then removed from the stack corresponding to that treatment).

Recall that $t$ is the number of periods elapsed divided by $n$.
Let 
\[
q_{a}(t):=\frac{1}{n}\sum_{j=1}^{\left\lfloor nt\right\rfloor }\mathbb{I}(A_{j}=a),
\]
and take $\mathcal{F}_{t}$ to be the $\sigma$-algebra generated
by
\begin{equation}
\xi_{t}:=\left\{ \{A_{j}\}_{j=1}^{\left\lfloor nt\right\rfloor },\{Y_{i}^{(1)}\}_{i=1}^{\left\lfloor nq_{1}(t)\right\rfloor },\{Y_{i}^{(0)}\}_{i=1}^{\left\lfloor nq_{0}(t)\right\rfloor }\right\} ,\label{eq:state=000020variables}
\end{equation}
the set of all actions and rewards until period $nt$. The sequence
of $\sigma$-algebras, $\{\mathcal{F}_{t}\}_{t\in\mathcal{T}_{n}}$,
where $\mathcal{T}_{n}:=\{1/n,2/n,\dots\}$, constitutes a filtration.
We require $\pi_{nt}(\cdot)$ to be $\mathcal{F}_{t-1/n}$ measurable,
the stopping time, $\tau$, to be $\mathcal{F}_{t-1/n}$ measurable,
and the implementation rule, $\delta$, to be $\mathcal{F}_{\tau}$
measurable. The set of all decision rules $\bm{d}\equiv(\{\pi_{nt}\}_{t\in\mathcal{T}_{n}},\tau,\delta)$
satisfying these requirements is denoted by $\mathcal{D}_{n}$. As
unbounded stopping times pose technical challenges, we generally work
with $\mathcal{D}_{n,T}\equiv\left\{ \bm{d}\in\mathcal{D}_{n}:\tau\le T\ \textrm{a.s}\right\} $,
the set of all decision rules with stopping times bounded by some
arbitrarily large, but finite, $T$. 

The mean outcomes under a parameter $\theta$ are denoted by $\mu_{a}(\theta):=\mathbb{E}_{P_{\theta}^{(a)}}[Y_{i}^{(a)}]$.
Following \citet{hirano2009asymptotics}, for each $a\in\{0,1\}$,
we consider local perturbations of the form $\{\theta_{0}^{(a)}+h_{a}/\sqrt{n};h_{a}\in\mathbb{R}^{d}\}$,
with $h_{a}$ unknown, around a reference parameter $\theta_{0}^{(a)}$.
As in that paper, $\theta_{0}^{(a)}$ is chosen such that $\mu_{1}(\theta_{0}^{(1)})=\mu_{0}(\theta_{0}^{(0)})=0$;
the last equality, which sets the quantities to $0$, is not necessary
and is simply a convenient re-centering. This choice of $\theta_{0}^{(a)}$
defines the hardest instance of the generalized Wald problem. When
$\mu_{1}(\theta_{0}^{(1)})\neq\mu_{0}(\theta_{0}^{(0)})$, determining
the best treatment is trivial under large $n$, and many decision
rules, including the one we propose here (in Section \ref{subsec:Attaining-the-bound-parametric}),
would achieve zero asymptotic regret. 

Let $P_{h}^{(a)}:=P_{\theta_{0}^{(a)}+h/\sqrt{n}}^{(a)}$ and take
$\mathbb{E}_{h}^{(a)}[\cdot]$ to be its corresponding expectation.
We assume $P_{\theta}^{(a)}$ is differentiable in quadratic mean
around $\theta_{0}^{(a)}$ with score functions $\psi_{a}(Y_{i})$
and information matrices $I_{a}:=\mathbb{E}_{0}^{(a)}[\psi_{a}\psi_{a}^{\intercal}]$.
For each $h\in\mathbb{R}^{d}$, denote
\[
\mu_{n,a}(h):=\mu_{a}(\theta_{0}^{(a)}+h/\sqrt{n})\approx\dot{\mu}_{a}^{\intercal}h/\sqrt{n},
\]
where $\dot{\mu}_{a}:=\nabla_{\theta}\mu_{a}(\theta_{0}^{(a)})$.
To reduce some notational overhead, we set $\theta_{0}^{(1)}=\theta_{0}^{(0)}=\theta_{0}$,
and also suppose that $\mu_{n,a}(h)=-\mu_{n,a}(-h)$ for all $h$.
The latter is always true asymptotically. Both simplifications can
be easily dispensed with, at the expense of some additional notation:
we emphasize that our results do not fundamentally require $\theta_{0}^{(1)},\theta_{0}^{(0)}$
to be the same or even have the same dimension.

\subsection{Bayes and minimax regret under fixed $n$}

Let $P_{n,h}^{(a)}$ denote the joint probability over ${\bf y}_{nT}^{(a)}:=\left\{ Y_{1}^{(a)},\dots,Y_{nT}^{(a)}\right\} $
- the largest possible (under $\tau\le T$) iid sequence of outcomes
that can be observed from treatment $a$ - when $Y^{(a)}\sim P_{h}^{(a)}$.
Define $\bm{h}:=(h_{1},h_{0})$, take $P_{n,\bm{h}}$ to be the joint
probability $P_{n,h_{1}}^{(1)}\times P_{n,h_{0}}^{(0)}$, and $\mathbb{E}_{n,\bm{h}}[\cdot]$
its corresponding expectation. The frequentist regret of decision
rule $\bm{d}$ is defined as 
\begin{align*}
V_{n}(\bm{d},\bm{h}) & =V_{n}\left(\bm{d},\left(\mu_{n,1}(h_{1}),\mu_{n,0}(h_{0})\right)\right)\\
 & :=\sqrt{n}\mathbb{E}_{n,\bm{h}}\left[\max\left\{ \mu_{n,1}(h_{1})-\mu_{n,0}(h_{0}),0\right\} -\left(\mu_{n,1}(h_{1})-\mu_{n,0}(h_{0})\right)\delta+\frac{c}{n^{3/2}}n\tau\right]\\
 & =\sqrt{n}\mathbb{E}_{n,\bm{h}}\left[\max\left\{ \mu_{n,1}(h_{1})-\mu_{n,0}(h_{0}),0\right\} -\left(\mu_{n,1}(h_{1})-\mu_{n,0}(h_{0})\right)\delta\right]+c\mathbb{E}_{n,\bm{h}}[\tau],
\end{align*}
where the multiplication by $\sqrt{n}$ in the second line of the
above equation is a normalization ensuring $V_{n}(\bm{d},\bm{h})$
converges to a non-trivial quantity.

Let $\nu$ denote a dominating measure over $\{P_{\theta}:\theta\in\Theta\}$,
and define $p_{\theta}:=dP_{\theta}/d\nu$. Also, take $M_{0}$ to
be some prior over $\bm{h}$, and $m_{0}$ its density with respect
to some other dominating measure $\nu_{1}$. By \citet{adusumilli2021risk},
the posterior density (wrt $\nu_{1}$), $p_{n}(\cdot\vert\mathcal{F}_{t})$,
of $\bm{h}$ depends only on ${\bf y}_{nq_{a}(t)}^{(a)}=\{Y_{i}^{(a)}\}_{i=1}^{\left\lfloor nq_{a}(t)\right\rfloor }$
for $a\in\{0,1\}$. Hence,
\begin{align}
p_{n}(\bm{h}\vert\mathcal{\mathcal{F}}_{t}) & =p_{n}\left(\bm{h}\vert{\bf y}_{nq_{1}(t)}^{(1)},{\bf y}_{nq_{0}(t)}^{(0)}\right)\nonumber \\
 & \propto\left\{ \prod_{i=1}^{\left\lfloor nq_{1}(t)\right\rfloor }p_{\theta_{0}+h_{1}/\sqrt{n}}^{(1)}(Y_{i}^{(1)})\right\} \left\{ \prod_{i=1}^{\left\lfloor nq_{0}(t)\right\rfloor }p_{\theta_{0}+h_{0}/\sqrt{n}}^{(0)}(Y_{i}^{(0)})\right\} m_{0}(\bm{h}).\label{eq:True=000020posterior-2}
\end{align}
The fixed $n$ Bayes regret of a decision $\bm{d}$ is given by $V_{n}(\bm{d},m_{0}):=\int V_{n}(\bm{d},\bm{h})dm_{0}(\bm{h})$. 

Following definition (\ref{eq:state=000020variables}), let $\xi_{\tau}$
denote the set of all actions and rewards generated over the course
of the experiment. From the form of $V_{n}(\bm{d},\bm{h})$, it is
clear that the Bayes optimal implementation rule is $\delta^{*}(\xi_{\tau})=\mathbb{I}\left\{ \mu_{n,1}(\xi_{\tau})\ge\mu_{n,0}(\xi_{\tau})\right\} $,
and the resulting Bayes regret at the terminal state is
\begin{equation}
\varpi_{n}(\xi_{\tau}):=\mu_{n}^{\max}(\xi_{\tau})-\max\left\{ \mu_{n,1}(\xi_{\tau}),\mu_{n,0}(\xi_{\tau})\right\} ,\label{eq:utility=000020function}
\end{equation}
where $\mu_{n,a}(\xi_{\tau}):=\mathbb{E}_{\bm{h}\vert\xi_{\tau}}[\mu_{n,a}(h_{a})]$
and $\mu_{n}^{\max}(\xi_{\tau}):=\mathbb{E}_{\bm{h}\vert\xi_{\tau}}[\max\{\mu_{n,1}(h_{1}),\mu_{n,0}(h_{0})\}]$.
We can thus associate each combination, $(\pi,\tau)$, of sampling
rules and stopping times with the distribution $\mathbb{P}_{\pi,\tau}$
that they induce over $(\varpi_{n}(\xi_{\tau}),\tau)$. Thus,
\[
V_{n}\left(\bm{d},m_{0}\right)=\mathbb{E}_{\pi,\tau}\left[\sqrt{n}\varpi_{n}(\xi_{\tau})+c\tau\right].
\]
For any given $T<\infty$, the minimal Bayes regret in the fixed $n$
setting is therefore
\[
V_{n,T}^{*}(m_{0})=\inf_{\bm{d}\in\mathcal{D}_{n,T}}\mathbb{E}_{\pi,\tau}\left[\sqrt{n}\varpi_{n}(\xi_{\tau})+c\tau\right].
\]

While our interest is in minimax regret, $V_{n,T}^{*}:=\inf_{\bm{d}\in\mathcal{D}_{n,T}}\sup_{\bm{h}}V_{n}(\bm{d},\bm{h})$,
the minimal Bayes regret is a useful theoretical device as it provides
a lower bound, $V_{n,T}^{*}\ge V_{n,T}^{*}(m_{0})$ for any prior
$m_{0}$. 

\subsection{Lower bound on minimax regret}

We impose the following assumptions (here, and in what follows, $\vert\cdot\vert$
denotes the Euclidean norm):

\begin{asm1} (i) The class $\{P_{\theta}^{(a)};\theta\in\mathbb{R}^{d}\}$
is differentiable in quadratic mean around $\theta_{0}$ for each
 $a\in\{0,1\}$. (ii) $\mathbb{E}_{0}^{(a)}[\exp\vert\psi_{a}(Y_{i}^{(a)})\vert]<\infty$
for $a\in\{0,1\}$. (iii) There exist $\dot{\mu}_{1},\dot{\mu}_{0}$
and $\epsilon_{n}^{(1)},\epsilon_{n}^{(0)}\to0$ s.t $\sqrt{n}\mu\left(P_{h}^{(a)}\right):=\sqrt{n}\mu_{n,a}(h)=\dot{\mu}_{a}^{\intercal}h+\epsilon_{n}^{(a)}\vert h\vert^{2}$
for each $a\in\{0,1\}$ and $h\in\mathbb{R}^{d}$.\end{asm1}

The assumptions are standard, with the only onerous requirement being
Assumption 1(ii), which requires the score function to have bounded
exponential moments. This is needed due to the proof techniques, which
are adapted from \citet{adusumilli2021risk}. 

Let $V^{*}$ denote the asymptotic minimax regret, defined as the
value of the minimax problem in (\ref{eq:def=000020of=000020minimax=000020risk}). 

\begin{thm} \label{Thm:=000020Parametric=000020families=000020lower=000020bound}Suppose
Assumptions 1(i)-(iii) hold. Then, 
\[
\sup_{\mathcal{J}}\lim_{T\to\infty}\liminf_{n\to\infty}\inf_{\bm{d}\in\mathcal{D}_{n,T}}\sup_{\bm{h}\in\mathcal{J}}V_{n}(\bm{d},\bm{h})\ge V^{*},
\]
where the outer supremum is taken over all finite subsets $\mathcal{J}$
of $\mathbb{R}^{d}\times\mathbb{R}^{d}$.\end{thm} 

It is straightforward to extend Theorem \ref{Thm:=000020Parametric=000020families=000020lower=000020bound}
to best arm identification. We omit the formal statement for brevity.
The proof proceeds as follows: Let $\sigma_{a}^{2}:=\dot{\mu}_{a}^{\intercal}I_{a}^{-1}\dot{\mu}_{a}$,
\[
h_{a}^{*}:=\frac{\sigma_{a}\Delta^{*}}{2\dot{\mu}_{a}^{\intercal}I_{a}^{-1}\dot{\mu}_{a}}I_{a}^{-1}\dot{\mu}_{a},
\]
and take $m_{0}^{*}$ to be the symmetric two-prior supported on $(h_{1}^{*},-h_{0}^{*})$
and $(-h_{1}^{*},h_{0}^{*}$). This is the parametric counterpart
to the least-favorable prior described in Theorem \ref{theorem=0000201}.
Clearly, there exist subsets $\mathcal{J}$ such that
\[
\inf_{\bm{d}\in\mathcal{D}_{n,T}}\sup_{\bm{h}\in\mathcal{J}}V_{n}(\bm{d},\bm{h})\ge\inf_{\bm{d}\in\mathcal{D}_{n,T}}V_{n}(\bm{d},m_{0}^{*}).
\]
In Appendix \ref{sec:Appendix:A}, we show
\begin{equation}
\lim_{T\to\infty}\lim_{n\to\infty}\inf_{\bm{d}\in\mathcal{D}_{n,T}}V_{n}(\bm{d},m_{0}^{*})=V^{*}.\label{eq:lower=000020bound=000020-=000020intermediate=000020result}
\end{equation}
To prove (\ref{eq:lower=000020bound=000020-=000020intermediate=000020result}),
we build on previous work in \citet{adusumilli2021risk}. Standard
techniques, such as asymptotic representation theorems \citep{van2000asymptotic},
are not easily applicable here due to the continuous time nature of
the problem. We instead employ a three step approach: First, we replace
$P_{n,\bm{h}}$ with a simpler family of measures whose likelihood
ratios (under different values of $\bm{h}$) are the same as those
under Gaussian distributions. Then, for this family, we write down
a HJB-Variational Inequality (HJB-VI) to characterize the optimal
value function under fixed $n$. PDE approximation arguments then
let us approximate the fixed-$n$ value function with that under continuous
time. The latter is shown to be $V^{*}$. 

The definition of asymptotic minimax risk used in Theorem \ref{Thm:=000020Parametric=000020families=000020lower=000020bound}
is standard, see, e.g., \citet[Theorem 8.11]{van2000asymptotic},
apart from the $\lim_{T\to\infty}$ operation. The theorem asserts
that $V^{*}$ is a lower bound on minimax regret under any bounded
stopping time. The bound $T$ can be arbitrarily large. Our proof
techniques require bounded stopping times as various approximation
results, e.g., the SLAN property (see, equation (\ref{eq:LAN}) in
Appendix \ref{sec:Appendix:A}), are only valid when the experiment is of bounded duration.\footnote{For any given $\hm{h}$, the dominated convergence theorem implies
$\lim_{T\to\infty}\inf_{\bm{d}\in\mathcal{D}_{n,T}}V_{n}(\bm{d},\bm{h})=\inf_{\bm{d}\in\mathcal{D}_{n}}V_{n}(\bm{d},\bm{h})$.
However, to allow $T=\infty$ in Theorem \ref{theorem=0000201}, we
need to show that this equality holds uniformly over $n$. In specific
instances, e.g., when the parametric family is Gaussian, this is indeed
the case, but we are not aware of any general results in this direction. } Nevertheless, we conjecture that there is no loss in setting $T=\infty$
in practice. 

\subsection{Attaining the bound\protect\label{subsec:Attaining-the-bound-parametric}}

We now describe a decision rule $\bm{d}_{n}=(\pi_{n},\tau_{n},\delta_{n})$
that is asymptotically minimax optimal. Let $\sigma_{a}^{2}=\dot{\mu}_{a}^{\intercal}I_{a}^{-1}\dot{\mu}_{a}$
for each $a$ and
\[
\rho_{n}(t):=\frac{x_{1}(t)}{\sigma_{1}}-\frac{x_{0}(t)}{\sigma_{0}},\ \textrm{where}\quad x_{a}(t):=\frac{\dot{\mu}_{a}^{\intercal}I_{a}^{-1}}{\sqrt{n}}\sum_{i=1}^{\left\lfloor nq_{a}(t)\right\rfloor }\psi_{a}(Y_{i}^{(a)}).
\]
Note that $x_{a}(t)$ is the efficient influence function process
for estimation of $\mu_{a}(\theta)$. We assume $\dot{\mu}_{a},I_{a},\sigma_{a}$
are known; but in practice, they should be replaced with consistent
estimates (from a vanishingly small initial sample) so that they do
not require knowledge of the reference parameter $\theta_{0}$. As
described in the next section, this can be done without affecting
the asymptotic results. 

Take $\pi_{n}$ to be any sampling rule such that 
\begin{equation}
\left|\frac{q_{a}(t)}{t}-\frac{\sigma_{a}}{\sigma_{1}+\sigma_{0}}\right|\le B\left\lfloor nt\right\rfloor ^{-b_{0}}\ \textrm{uniformly over bounded }t,\label{eq:requirement=000020on=000020pi_n}
\end{equation}
for some $B<\infty$ and $b_{0}>1/2$. To simplify matters, we suppose
that $\pi_{n}$ is deterministic, e.g., $\pi_{n,1}(t)=\mathbb{I}\left\{ q_{1}(t)\le t\sigma_{1}/(\sigma_{1}+\sigma_{0})\right\} $.
Fully randomized rules, $\pi_{n,1}(t)=\sigma_{1}/(\sigma_{0}+\sigma_{1})$,
do not satisfy the `fine-balance' condition (\ref{eq:requirement=000020on=000020pi_n})
and we indeed found them to perform poorly in simulations. We further
employ
\[
\tau_{n,T}=\inf\left\{ t:\left|\rho_{n}(t)\right|\ge\gamma^{*}\right\} \wedge T
\]
as the stopping time, and as the implementation rule, set $\delta_{n,T}=\mathbb{I}\left\{ \rho_{n}(\tau_{n,T})\ge0\right\} $.

Intuitively, $\bm{d}_{n,T}=(\pi_{n},\tau_{n,T},\delta_{n,T})$ is
the finite sample counterpart of the minimax optimal decision rule
$\bm{d}^{*}$ from Section \ref{sec:Minimax-regret}. The following
theorem shows that it is asymptotically minimax optimal in that it
attains the lower bound of Theorem \ref{Thm:=000020Parametric=000020families=000020lower=000020bound}.

\begin{thm} \label{Thm:=000020Parametric=000020-=000020attaining=000020the=000020bound}Suppose
Assumptions 1(i)-(iii) hold. Then, 
\[
\sup_{\mathcal{J}}\lim_{T\to\infty}\liminf_{n\to\infty}\sup_{\bm{h}\in\mathcal{J}}V_{n}(\bm{d}_{n,T},\bm{h})=V^{*},
\]
where the outer supremum is taken over all finite subsets $\mathcal{J}$
of $\mathbb{R}^{d}\times\mathbb{R}^{d}$.\end{thm}

An important implication of Theorem \ref{Thm:=000020Parametric=000020-=000020attaining=000020the=000020bound}
is that the minimax optimal decision rule only involves one state
variable, $\rho_{n}(t)$. This is even though the state space in principle
includes all the past observations until period $i$, for a total
of at least $2i$ variables. The theorem thus provides a major reduction
in dimension.

\subsection{Unknown variances\protect\label{subsec:Unknown-variances}}

Replacing $\sigma_{1},\sigma_{0}$ (and other population quantities)
with consistent estimates has no effect on asymptotic regret. We suggest
two approaches to attain the minimax lower bounds when these parameters
are unknown.

The first approach uses `forced exploration' (see, e.g., \citealp[Chapter 33, Note 7]{lattimore2020bandit}):
we set $\pi_{n}^{*}(t)=1/2$ for the first $\bar{n}=n^{a}$ observations,
where $a\in(0,1)$. This corresponds to a time duration of $\bar{t}=n^{a-1}$.
We use the data from these periods to obtain consistent estimates,
$\hat{\sigma}_{1}^{2},\hat{\sigma}_{0}^{2}$ of $\sigma_{1}^{2},\sigma_{0}^{2}$.
From $\bar{t}$ onwards, we apply the minimax optimal rule $\bm{d}_{n,T}$
after plugging-in $\hat{\sigma}_{1},\hat{\sigma}_{0}$ in place of
$\sigma_{1},\sigma_{0}$. Note that when applying $\bm{d}_{n,T}$,
we should start $x_{1}(\cdot),x_{0}(\cdot)$ from their values at
$\bar{t}$ to ensure the information accrued before $\bar{t}$ is
also taken into account. This strategy is asymptotically minimax optimal
for any $a\in(0,1)$. 

Our second suggestion is to place a prior on $\sigma_{1},\sigma_{0}$,
and continuously revise their values using the posterior means. We
recommend using an inverse-gamma prior and computing the posterior
by treating the scores $\psi_{a}(Y_{i}^{(a)})$ as Gaussian (which
is justified asymptotically). This approach has the advantage of not
requiring any tuning parameters. 

Admittedly, both proposals treat estimation of $\sigma_{1},\sigma_{0}$
as separate and somewhat less critical than the estimation of the
population mean parameters. However, this merely reflects the significant
asymmetry in the complexity of estimating these parameters in the
continuous time setting. As noted in Section \ref{sec:Diffusion-asymptotics-and},
$\sigma_{1},\sigma_{0}$ can be learnt instantly in continuous time
from the quadratic variations of $x_{1}(t),x_{0}(t)$. Conversely,
running the sequential experiment for a brief period would only marginally
update the prior over $\bm{\mu}$. 

Furthermore, in the finite $n$ setting, it is important to recognize
that $\theta^{(a)}$ characterizes the entire distribution of $Y^{(a)}$,
including its mean and variance.\textbf{ }The quantities\textbf{ $\sigma_{1},\sigma_{0}$
}do not represent the variances of the underlying probability distributions
- which are subject to change anyway under the local sequence $\theta_{0}^{(a)}+h_{a}/\sqrt{n}$
- but are rather the information matrices evaluated at the reference
parameters\textbf{ $\theta_{0}^{(1)},\theta_{0}^{(0)}$.} Conceptually,
under local asymptotics, these reference parameters are assumed to
be known in advance. While one would aim, in practice, to construct
procedures that adapt to or are invariant to these quantities, the
impact of estimating them cannot be accounted for in the local asymptotic
framework itself. This is not to diminish the importance of efficiently
estimating $\sigma_{1},\sigma_{0}$; rather, the issue lies outside
the scope of first-order asymptotic theory, which is just too coarse
an approximation for this purpose. Addressing this would require employing
higher-order asymptotics. 

\section{Numerical illustration\protect\label{sec:Numerical-illustration}}

A/B testing is commonly used in online platforms for optimizing websites.
Consequently, to assess the finite sample performance of our proposed
policies, we run a Monte-Carlo simulation calibrated to a realistic
example of such an A/B test. Suppose there are two candidate website
layouts, with exit rates $\gamma_{0},\gamma_{1}$, and we want to
run an A/B test to determine the one with the lowest exit rate.\footnote{The exit rate is defined as the fraction of viewers of a webpage who
exit from the website it is part of (i.e., without viewing other pages
in that website).} The outcomes are binary, $Y^{(a)}\sim\textrm{Bernoulli}(\gamma_{a})$.
This is a parametric setting with score functions $\psi_{a}(Y_{i}^{(a)})=Y_{i}^{(a)}$.
We calibrate $\gamma_{0}=0.4$, which is a typical value for an exit
rate. The cost of experimentation is normalized to $c=1$ and we consider
various values of $n$, corresponding to different `population sizes'
(recall that the benefit during implementation is scaled as $n^{3/2}\gamma_{a}$).
We then set $\gamma_{1}=\gamma_{0}+\Delta/\sqrt{n}$, and describe
the results under varying $\Delta$. Local asymptotics provide a good
approximation in practice because raw performance gains are generally
small - typically, $\vert\gamma_{1}-\gamma_{0}\vert$ is of the order
0.05 or less (see, e.g., \citealp{deng2013improving}) - but these
gains can translate into large profits when applied at scale, i.e.,
when $n$ is large.

Since $\sigma_{a}=\sqrt{\gamma_{a}(1-\gamma_{a})}$ is unknown, we
employ `forced sampling' with $\bar{n}=\max(50,0.05n)$, i.e., using
about 5\% of the sample, to estimate $\sigma_{1},\sigma_{0}$. Note
that the asymptotically optimal sampling rule is always $1/2$ in
the Bernoulli setting, so forced sampling is in fact asymptotically
costless. We also experimented with a beta prior to continuously update
$\sigma_{a}$, but found the results to be somewhat inferior, (see
Appendix C for details). Figure \ref{fig:Minimax=000020risk=000020Bernoulli},
Panel A plots the finite sample frequentist regret profiles of our
policy rules $\bm{d}_{n}:=\bm{d}_{n,\infty}$ (with $T=\infty$) for
various values of $n$, along with that of the minimax optimal policy
$\bm{d}^{*}$ under the diffusion regime; the regret profile of the
latter is derived analytically in Lemma \ref{Lemma=0000203}. Diffusion
asymptotics provide a very good approximation to the finite sample
properties of $\bm{d}_{n}$, even for such relatively small values
of $n$ as $n=1000$. In practice, A/B tests often involve tens, even
hundreds, of thousands of observations. The max-regret of $\bm{d}_{n}$
is also very close to the asymptotic lower bound $V^{*}$ (the max-regret
of $\bm{d}^{*}$).

Figure \ref{fig:Minimax=000020risk=000020Bernoulli}, Panel B displays
some summary statistics for the Bayes regret of $\bm{d}_{n}$ under
the least-favorable prior, $p_{\Delta^{*}}$. The regret distribution
is positively skewed and heavy tailed. The finite sample Bayes regret
is again very close to $V^{*}$. 

Appendix C reports additional simulation results using Gaussian outcomes.
\begin{figure}
\includegraphics[height=5cm]{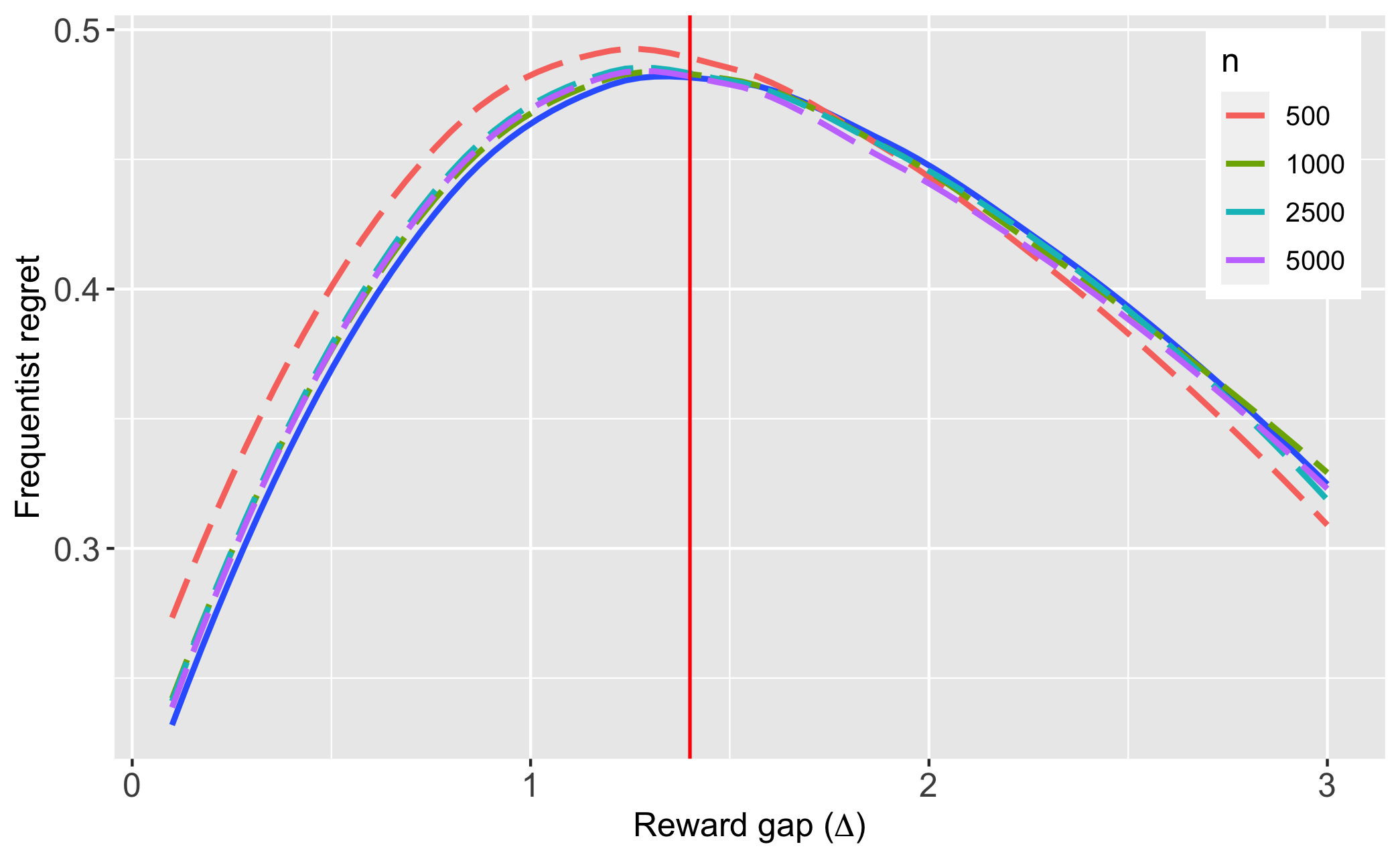}~\includegraphics[height=5cm]{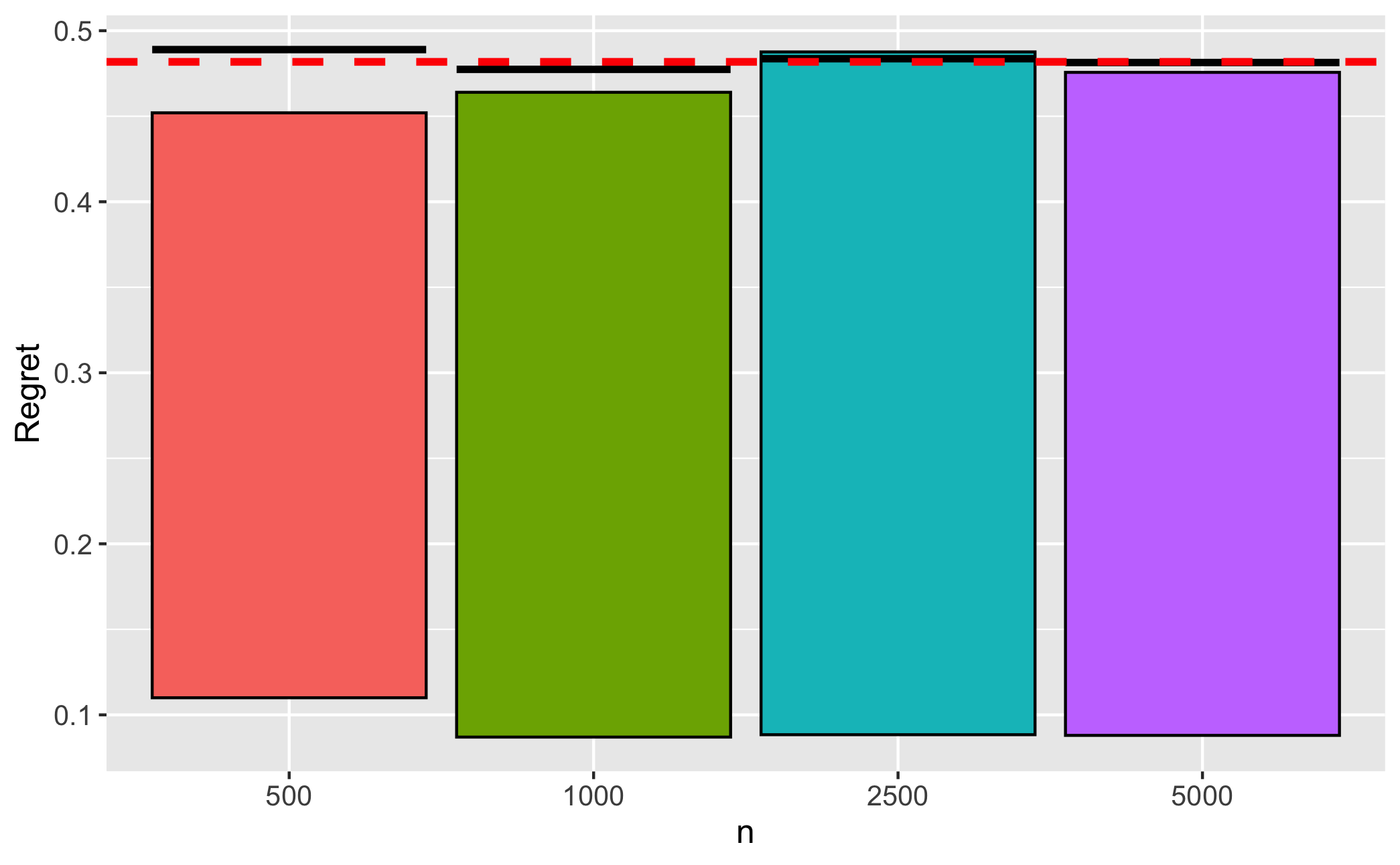}~

\begin{tabular}{>{\centering}p{7.5cm}>{\centering}p{7.5cm}}
{\scriptsize A: Frequentist regret profiles} & {\scriptsize$\ $B: Performance under least-favorable prior}\tabularnewline
\end{tabular}

\vspace{0.5em}
\begin{raggedright}
{\scriptsize Note: The solid curve in Panel A is the regret profile
of $\bm{d}^{*}$; the vertical line denotes $\Delta^{*}$. We only
plot the results for $\Delta>0$ as the values are close to symmetric.
The dashed line in Panel B is $V^{*}$, the asymptotic minimax regret.
The solid horizontal lines denote the Bayes regret in finite samples,
under the least-favorable prior. The bars describe the interquartile
range of regret.}{\scriptsize\par}
\par\end{raggedright}
\caption{Finite sample performance of $\bm{d}_{n}$\protect\label{fig:Minimax=000020risk=000020Bernoulli}}
\end{figure}

\section{Variations and extensions\protect\label{sec:Variations-and-extensions}}

We now consider various modifications of the basic setup and analyze
if, and how, the optimal decisions change. Appendix G discusses extensions
to multiple treatments. 

\subsection{Batching}

In practice, it may be that data is collected in batches instead of
one at a time, and the DM can only make decisions after processing
each batch. Let $B_{n}$ denote the number of observations considered
in each batch. In the context of Section \ref{sec:Formal-properties},
this corresponds to a time duration of $B_{n}/n$. An analysis of
its proof shows that Theorem \ref{Thm:=000020Parametric=000020families=000020lower=000020bound}
continues to hold as long as $B_{n}/n\to0$. Thus, $\bm{d}_{n,T}$
remains asymptotically minimax optimal in this scenario. 

Even for $B_{n}/n\to m\in(0,1)$, the optimal decision rules remain
broadly unchanged. Asymptotically, we have equivalence to Gaussian
experiments, so we can analyze batched experiments under the diffusion
framework by imagining that the stopping time is only allowed to take
on discrete values $\{0,1/m,2/m,\dots\}$. It is then clear from the
discussion in Section \ref{subsec:Intuition-behind-Theorem} that
the optimal sampling and implementation rules remain unchanged. The
discrete nature of the setting makes determining the optimal stopping
rule difficult, but it is easy to show that the decision rule $(\pi^{*},\tau_{m}^{*},\delta^{\tau_{m}^{*}})$,
where 
\[
\tau_{m}^{*}:=\inf\left\{ t\in\{0,1/m,2/m,\dots\}:\left|\frac{x_{1}(t)}{\sigma_{1}}-\frac{x_{0}(t)}{\sigma_{0}}\right|\ge\gamma^{*}\right\} 
\]
and $\delta^{\tau_{m}^{*}}:=\mathbb{I}\left\{ \sigma_{1}^{-1}x_{1}(\tau_{m}^{*})-\sigma_{0}^{-1}x_{0}(\tau_{m}^{*})\ge0\right\} $,
while not being exactly optimal, has a minimax regret that is arbitrarily
close to $V^{*}$ for large enough $m$ (note that no batched experiment
can attain a minimax regret that is lower than $V^{*}$). 

\subsection{Alternative cost functions\protect\label{subsec:Alternative-cost-functions-extension}}

All our results so far were derived under constant sampling costs.
The same techniques apply to other types of flow costs as long as
these depend only on $\rho(t):=\sigma_{1}^{-1}x_{1}(t)-\sigma_{0}^{-1}x_{0}(t)$.
In particular, suppose that the frequentist regret is given by 
\[
V\left(\bm{d},\bm{\mu}\right)=\mathbb{E}_{\bm{d}\vert\bm{\mu}}\left[\max\{\mu_{1}-\mu_{0},0\}-(\mu_{1}-\mu_{0})\delta+\int_{0}^{\tau}c(\rho(t))dt\right],
\]
where $c(z)$ is the flow cost of experimentation when $\rho(t)=z$.
We require $c(\cdot)$ to be (i) positive, (ii) bounded away from
$0$, i.e., $\inf_{z}c(z)\ge\underline{c}>0$, and (iii) symmetric,
i.e., $c(z)=c(-z)$. By (\ref{eq:evolution=000020of=000020rho}),
$(\sigma_{1}+\sigma_{0})\rho(t)/t$ is an estimate of the treatment
effect $\mu_{1}-\mu_{0}$, so the above allows for situations in which
sampling costs depend on the magnitude of the estimated treatment
effects. While we are not aware of any real world examples of such
costs, they could arise if there is feedback between the observations
and sampling costs, e.g., if it is harder to find subjects for experimentation
when the treatment effect estimates are higher. When there are only
two states, the `ex-ante' entropy cost of \citet{sims2003implications}
is also equivalent to a specific flow cost of the form $c(\cdot)$
above, see \citet{morris2019wald}.\footnote{However, we are not aware of any extension of this result to continuous
states.} 

For the above class of cost functions, we show in Appendix D that
the minimax optimal decision rule $\bm{d}^{*}$ and the least-favorable
prior $p_{\Delta}^{*}$ have the same form as in Theorem \ref{theorem=0000201},
but the values of $\gamma^{*},\Delta^{*}$ are different and need
to be calculated by solving the minimax problem 
\[
\min_{\gamma}\max_{\Delta}\left\{ \left(\frac{\sigma_{1}+\sigma_{0}}{2}\right)\frac{\left(1-e^{-\Delta\gamma}\right)\Delta}{e^{\Delta\gamma}-e^{-\Delta\gamma}}+\frac{\left(e^{\Delta\gamma}-1\right)\zeta_{\Delta}(\gamma)+\left(1-e^{-\Delta\gamma}\right)\zeta_{\Delta}(-\gamma)}{e^{\Delta\gamma}-e^{-\Delta\gamma}}\right\} ,
\]
where 
\[
\zeta_{\Delta}(x):=2\int_{0}^{x}\int_{0}^{y}e^{\Delta(z-y)}c(z)dzdy.
\]

Beyond this class of sampling costs, however, it is easy to conceive
of scenarios in which the optimal decision rule differs markedly from
the one we obtain here. For instance, Neyman allocation would no longer
be the optimal sampling rule if the costs for sampling each treatment
were different. Alternatively, if $c(\cdot)$ were to depend on $t$,
the optimal stopping time could have a very different form. The analysis
of these cost functions is not covered by the present techniques.

\subsection{Non-parametric outcomes\protect\label{subsec:Beyond-mean-regret}}

In Appendix E, we extend Theorems \ref{Thm:=000020Parametric=000020families=000020lower=000020bound}
and \ref{Thm:=000020Parametric=000020-=000020attaining=000020the=000020bound}
to the non-parametric setting, where there is no a-priori information
about the distributions $P^{(1)},P^{(0)}$ of $Y_{i}^{(1)}$ and $Y_{i}^{(0)}.$
The minimax optimal rule retains the same form as in Section \ref{subsec:Attaining-the-bound-parametric},
but with $x_{a}(t)$ now defined as $n^{-1/2}\sum_{i=1}^{\left\lfloor nq_{a}(t)\right\rfloor }Y_{i}^{(a)}$,
and $\sigma_{a}^{2}$ as the variance of $Y_{i}^{(a)}$ at some reference
distribution $P_{0}^{(a)}$ (as in Section \ref{sec:Formal-properties},
$P_{0}^{(1)},P_{0}^{(0)}$ are to be chosen such that $\mathbb{E}_{P_{0}^{(1)}}[Y_{i}^{(1)}]=\mathbb{E}_{P_{0}^{(0)}}[Y_{i}^{(0)}]$).
One can obtain these same results by simply assuming the outcomes
to be Gaussian.

The above results can also be extended to different regret measures.
Specifically, instead of $\mu(\cdot)$ denoting the mean functional
in the definition of regret $\max\{\mu(P^{(1)})-\mu(P^{(0)}),0\}-(\mu(P^{(1)})-\mu(P^{(0)}))\delta+c\tau$,
it can denote other functionals of the outcome distribution in the
implementation phase (we still need costs to be linear and additively
separable). For instance, $\mu(\cdot)$ could be the quantile function.
In Appendix E.4, we show that the decision rule $\bm{d}_{n.T}$ from
Section \ref{subsec:Attaining-the-bound-parametric} is still minimax
optimal if we just redefine $x_{a}(t)$ to now be the efficient influence
function process $n^{-1/2}\sum_{i=1}^{\left\lfloor nq_{a}(t)\right\rfloor }\psi_{a}(Y_{i}^{(a)})$,
where $\psi_{a}(\cdot)$ is the efficient influence function corresponding
to $\mu(P^{(a)})$.

\section{Conclusion}

This paper proposes a minimax-regret optimal procedure for determining
the best treatment when sampling is costly. The optimal sampling rule
is simply the Neyman allocation, while the optimal stopping rule advises
that the experiment be terminated when the average difference in outcomes
multiplied by the number of observations exceeds a specific threshold.
While these rules were derived under diffusion asymptotics, it is
shown that finite sample counterparts of these rules remain optimal
under both parametric and non-parametric regimes. The form of these
rules is robust to a number of different variations of the original
problem, e.g., under batching, different cost functions etc. Given
the simple nature of these rules, and the potential for large efficiency
gains (requiring, on average, 40\% fewer observations than standard
approaches), we believe they hold a lot of promise for practical use. 

\subsubsection*{Data availability statement}

The data and code underlying this article are available in Zenodo,
at \href{https://doi.org/10.5281/zenodo.14792035}{https://doi.org/10.5281/zenodo.14792035}.

\bibliographystyle{econ-econometrica}
\bibliography{Sequential_stopping}

\newpage{}

\appendix

\section{Proofs\protect\label{sec:Appendix:A}}

\subsection{Proof of Theorem \ref{theorem=0000201}}

The proof makes use of the following lemmas:

\begin{lem} \label{Lemma=0000201} Suppose nature sets $p_{0}$ to
be a symmetric two-point prior supported on $(\sigma_{1}\Delta/2,-\sigma_{0}\Delta/2),(-\sigma_{1}\Delta/2,\sigma_{0}\Delta/2$).
Then, the decision $d(\Delta)=(\pi^{*},\tau_{\gamma(\Delta)},\delta^{*})$,
where $\gamma(\Delta)$ is defined in (\ref{eq:definition=000020of=000020gamma(delta)}),
is a best response by the DM.\end{lem} 
\begin{proof}
The prior is an indifference-inducing one, so by the argument given
in Section \ref{subsec:Intuition-behind-Theorem}, the DM is indifferent
between any sampling rule $\pi$. Thus, $\pi_{a}^{*}(t)=\sigma_{a}/(\sigma_{1}+\sigma_{0})$
is a best-response to this prior. Also, the prior is symmetric with
$m_{0}=1/2$, so by (\ref{eq:allocation=000020rule}) and (\ref{eq:LR=000020process=000020indiffierence}),
the Bayes optimal implementation rule, for any given given stopping
time $\tau$, is
\begin{align*}
\delta^{\tau} & =\mathbb{I}\left\{ \ln\varphi^{\pi^{*}}(\tau)\ge0\right\} =\mathbb{I}\left\{ \frac{x_{1}(\tau)}{\sigma_{1}}-\frac{x_{0}(\tau)}{\sigma_{0}}\ge0\right\} .
\end{align*}

It remains to compute the Bayes optimal stopping time. Let $\lambda=1$
denote the event $\bm{\mu}=(\sigma_{1}\Delta/2,-\sigma_{0}\Delta/2)$,
with $\lambda=0$ otherwise. The discussion in Section \ref{subsec:Intuition-behind-Theorem}
implies that, conditional on $\lambda$, the distribution of the likelihood
ratio process $\varphi^{\pi}(t)$ does not depend on $\pi$ and evolves
as 
\begin{equation}
d\ln\varphi(t)=(2\lambda-1)\frac{\Delta^{2}}{2}dt+\Delta d\tilde{W}(t),\label{eq:LR=000020process=000020evolution}
\end{equation}
where $\tilde{W}(\cdot)$ is a one-dimensional Wiener process. By
a similar argument as in \citet[Section 4.2.1]{shiryaev2007optimal},
this in turn implies that the posterior probability $m^{\pi}(t):=\mathbb{P}^{\pi}(\lambda=1\vert\mathcal{F}_{t})$
is also independent of $\pi$ and evolves as\footnote{\citet[Section 4.2.1]{shiryaev2007optimal} analyzes Bayesian updating
under binary states, but the setup and notation are slightly different
from here. However, the likelihood ratio (LR) processes are the same
in both cases, making the problems, and the derivation of (\ref{eq:belief=000020evolution}),
equivalent. Specifically, in \citet[Equation 4.53]{shiryaev2007optimal},
the LR process evolves as $d\ln\varphi(t)=\frac{r}{\sigma^{2}}(d\xi(t)-\frac{r}{2}dt)$
where $d\xi(t):=r\theta dt+\sigma d\tilde{W}(t)$. Here, $r,\sigma\in\mathbb{R}^{+}$
are known, and $\theta\in\{0,1\}$ denotes the unknown binary state
of the world. Equating $r,\sigma,\theta$ with $\Delta,1,\lambda$
then gives the same LR process as (\ref{eq:LR=000020process=000020evolution}). } 
\begin{equation}
dm(t)=\Delta m(t)(1-m(t))d\tilde{W}(t).\label{eq:belief=000020evolution}
\end{equation}
Therefore, by (\ref{eq:Bayes=000020optimal=000020stopping=000020rule})
the optimal stopping time also does not depend on $\pi$ and is given
by
\begin{align}
\tau(\Delta) & =\inf_{\tau\in\mathcal{T}}\mathbb{E}\left[\varpi(m(\tau))+c\tau\right],\ \textrm{where}\label{eq:pf=000020thm=0000201:=000020objective=000020fn=000020for=000020stopping=000020rule}\\
\varpi(m) & :=\frac{(\sigma_{1}+\sigma_{0})}{2}\Delta\min\left\{ m,1-m\right\} .
\end{align}

Inspection of the objective function in (\ref{eq:pf=000020thm=0000201:=000020objective=000020fn=000020for=000020stopping=000020rule})
shows that this is exactly the same objective as in the Bayesian hypothesis
testing problem, analyzed previously by \citet{arrow1949bayes} and
\citet{morris2019wald}. We follow the analysis of the latter paper.
\citet{morris2019wald} show that instead of choosing the stopping
time $\tau$, it is equivalent to imagine that the DM chooses a probability
distribution $G$ over the posterior beliefs $m(\tau)$ at an `ex-ante'
cost 
\begin{align*}
c(G) & =\frac{2c}{\Delta^{2}}\int(1-2m)\ln\frac{1-m}{m}dG(m),
\end{align*}
subject to the constraint $\int mdG(m)=m_{0}=1/2$. The precise form
of $c(G)$ is due to \citet[Proposition 3]{morris2019wald}. Under
the distribution $G$, the expected regret, exclusive of sampling
costs, for the DM is 
\[
\int\varpi(m)dG(m)=\frac{(\sigma_{1}+\sigma_{0})}{2}\Delta\int\min\{m,1-m\}dG(m).
\]
Hence, the stopping time, $\tau$, that solves (\ref{eq:pf=000020thm=0000201:=000020objective=000020fn=000020for=000020stopping=000020rule})
is the one that induces the distribution $G^{*}$, defined as
\begin{align*}
G^{*} & =\argmin_{G:\int mdG(m)=\frac{1}{2}}\left\{ c(G)+\int\varpi(m)dG(m)\right\} \\
 & =\argmin_{G:\int mdG(m)=\frac{1}{2}}\int f(m)dG(m),
\end{align*}
where 
\[
f(m):=\frac{2c}{\Delta^{2}}(1-2m)\ln\frac{1-m}{m}+\frac{(\sigma_{1}+\sigma_{0})}{2}\Delta\min\{m,1-m\}.
\]
Clearly, $f(\cdot)$ is strictly convex on $[0,1/2]$ and $f(m)=f(1-m)$.
Hence, setting 
\[
\alpha(\Delta):=\argmin_{\alpha\in\left[0,\frac{1}{2}\right]}\left\{ \frac{(\sigma_{1}+\sigma_{0})}{2}\Delta\alpha+\frac{2c}{\Delta^{2}}(1-2\alpha)\ln\frac{1-\alpha}{\alpha}\right\} ,
\]
it is easy to see that $G^{*}$ is a two-point distribution, supported
on $\alpha(\Delta),1-\alpha(\Delta)$ with equal probability $1/2$. 

We now claim that this distribution is induced by the stopping time
$\tau_{\gamma(\Delta)}$, where
\begin{equation}
\gamma(\Delta):=\frac{1}{\Delta}\ln\frac{1-\alpha(\Delta)}{\alpha(\Delta)}.\label{eq:definition=000020of=000020gamma(delta)}
\end{equation}
To this end, observe that by (\ref{eq:belief=000020process}) and
(\ref{eq:LR=000020evolution}), 
\[
m(t)=\frac{\exp\left\{ \Delta\left(\frac{x_{1}(t)}{\sigma_{1}}-\frac{x_{0}(t)}{\sigma_{0}}\right)\right\} }{1+\exp\left\{ \Delta\left(\frac{x_{1}(t)}{\sigma_{1}}-\frac{x_{0}(t)}{\sigma_{0}}\right)\right\} }.
\]
We can then write $\tau_{\gamma(\Delta)}$ in terms of $m(t)$ as
\[
\tau_{\gamma(\Delta)}=\inf\left\{ t:m(t)\notin[\alpha(\Delta),1-\alpha(\Delta)]\right\} .
\]
This immediately implies that the support points of $m(\tau_{\gamma(\Delta)})$
are $\alpha(\Delta)$, $1-\alpha(\Delta)$; also,
\[
\mathbb{E}[m(\tau_{\gamma(\Delta)})]=\left(1-\alpha(\Delta)\right)\mathbb{P}\left(m(\tau_{\gamma(\Delta)})=1-\alpha(\Delta)\right)+\alpha(\Delta)\mathbb{P}\left(m(\tau_{\gamma(\Delta)})=\alpha(\Delta)\right).
\]
But $m(\cdot)$ is a martingale, so Doob's optional stopping theorem
implies $\mathbb{E}[m(\tau_{\gamma(\Delta)})]=\mathbb{E}[m(0)]=1/2$.
Equating the two expressions for $\mathbb{E}[m(\tau_{\gamma(\Delta)})]$
gives 
\[
\mathbb{P}\left(m(\tau_{\gamma(\Delta)})=1-\alpha(\Delta)\right)=\mathbb{P}\left(m(\tau_{\gamma(\Delta)})=\alpha(\Delta)\right)=1/2.
\]
Thus, $m(\tau_{\gamma(\Delta)})$ is distributed as $G^{*}$, and
the stopping time $\tau_{\gamma(\Delta)}$ is therefore the best response
to nature's prior.
\end{proof}
\begin{lem} \label{Lemma=0000202} Suppose $\bm{\mu}$ is such that
$\vert\mu_{1}-\mu_{0}\vert=\frac{\sigma_{1}+\sigma_{0}}{2}\Delta$.
Then, for any $\gamma,\Delta>0$, 
\[
V\left(\tilde{\bm{d}}_{\gamma},\bm{\mu}\right)=\frac{(\sigma_{1}+\sigma_{0})}{2}\Delta\frac{1-e^{-\Delta\gamma}}{e^{\Delta\gamma}-e^{-\Delta\gamma}}+\frac{2c\gamma}{\Delta}\frac{e^{\Delta\gamma}+e^{-\Delta\gamma}-2}{e^{\Delta\gamma}-e^{-\Delta\gamma}}.
\]
Thus, the frequentist regret of $\tilde{\bm{d}}_{\gamma}$ depends
on $\bm{\mu}$ only through $\vert\mu_{1}-\mu_{0}\vert$. \end{lem} 
\begin{proof}
Suppose that $\mu_{1}>\mu_{0}$. Define 
\[
\xi(t):=\left(\frac{x_{1}(t)}{\sigma_{1}}-\frac{x_{0}(t)}{\sigma_{0}}\right)\cdot\Delta.
\]
Note that under $\tilde{\bm{d}}_{\gamma}$ and $\bm{\mu}$, 
\[
\frac{x_{1}(t)}{\sigma_{1}}-\frac{x_{0}(t)}{\sigma_{0}}=\frac{\Delta}{2}t+\tilde{W}(t),
\]
where $\tilde{W}(\cdot)$ is one-dimensional Brownian motion. Hence
$\xi(t)=\frac{\Delta^{2}}{2}t+\Delta\tilde{W}(t).$ We can write the
stopping time $\tau_{\gamma}$ in terms of $\xi(t)$ as 
\begin{align*}
\tau_{\gamma} & =\inf\left\{ t:\left|\frac{x_{1}(t)}{\sigma_{1}}-\frac{x_{0}(t)}{\sigma_{0}}\right|\ge\gamma\right\} =\inf\left\{ t:\vert\xi(t)\vert\ge\Delta\gamma\right\} ,
\end{align*}
and the implementation rule as $\delta^{\tau_{\gamma}}=\mathbb{I}\left\{ \xi(\tau_{\gamma})\ge0\right\} =\mathbb{I}\left\{ \xi(\tau_{\gamma})=\Delta\gamma\right\} .$

Now, noting the form of $\xi(t)$, we can apply similar arguments
as in \citet[Section 4.2, Lemma 5]{shiryaev2007optimal}, to show
that
\[
\mathbb{E}\left[\tau_{\gamma}\vert\bm{\mu}\right]=\frac{2}{\Delta^{2}}\frac{\Delta\gamma\left(e^{\Delta\gamma}+e^{-\Delta\gamma}-2\right)}{e^{\Delta\gamma}-e^{-\Delta\gamma}}.
\]
Furthermore, following \citet[Section 4.2, Lemma 4]{shiryaev2007optimal},
we also have 
\[
\mathbb{P}(\delta^{\tau_{\gamma}}=0\vert\bm{\mu})=\mathbb{P}(\xi(\tau_{\gamma})=-\Delta\gamma\vert\bm{\mu})=\frac{1-e^{-\Delta\gamma}}{e^{\Delta\gamma}-e^{-\Delta\gamma}}.
\]
Hence, the frequentist regret is given by 
\begin{align*}
V\left(\tilde{\bm{d}}_{\gamma},\bm{\mu}\right) & =\frac{\sigma_{1}+\sigma_{0}}{2}\Delta\mathbb{P}(\delta^{\tau_{\gamma}}=0\vert\bm{\mu})+c\mathbb{E}\left[\tau_{\gamma}\vert\bm{\mu}\right]\\
 & =\frac{(\sigma_{1}+\sigma_{0})}{2}\Delta\frac{1-e^{-\Delta\gamma}}{e^{\Delta\gamma}-e^{-\Delta\gamma}}+\frac{2c\gamma}{\Delta}\frac{e^{\Delta\gamma}+e^{-\Delta\gamma}-2}{e^{\Delta\gamma}-e^{-\Delta\gamma}}.
\end{align*}

While the above was shown under $\mu_{1}>\mu_{0}$, an analogous argument
under $\mu_{1}<\mu_{0}$ gives the same expression for $V\left(\tilde{\bm{d}}_{\gamma},\bm{\mu}\right)$. 
\end{proof}
\begin{lem} \label{Lemma=0000203} Consider a two-player zero-sum
game in which nature chooses $\Delta\in\mathbb{R}^{+}$ indexing the
indifference prior $p_{\Delta}$ and the DM chooses $\gamma\in\mathbb{R}^{+}$
indexing the decision rule $\bm{d}_{\gamma}=(\pi^{*},\tau_{\gamma},\delta^{\tau_{\gamma}})$.
There exists a unique Nash equilibrium to this game at $\Delta^{*}=\eta\Delta_{0}^{*}$
and $\gamma^{*}=\eta^{-1}\gamma_{0}^{*}$, where $\eta,\Delta_{0}^{*},\gamma_{0}^{*}$
are defined in Section \ref{sec:Minimax-regret}. \end{lem} 
\begin{proof}
By Lemma \ref{Lemma=0000202}, the frequentist regret under a given
choice of $\Delta:=2\vert\mu_{1}-\mu_{0}\vert/(\sigma_{1}+\sigma_{0})$
and $\gamma$ is given by $\frac{(\sigma_{1}+\sigma_{0})}{2}R(\gamma,\Delta)$,
where
\[
R(\gamma,\Delta):=\Delta\frac{1-e^{-\Delta\gamma}}{e^{\Delta\gamma}-e^{-\Delta\gamma}}+\frac{2\eta^{3}\gamma}{\Delta}\frac{e^{\Delta\gamma}+e^{-\Delta\gamma}-2}{e^{\Delta\gamma}-e^{-\Delta\gamma}}.
\]
Lemma \ref{Lemma=0000202} further implies that the frequentist regret
$V(\bm{d}^{*},\bm{\mu})$ depends on $\bm{\mu}$ only through $\Delta$.
Therefore, the frequentist regret under both support points of $p_{\Delta}$
must be the same. Hence, the Bayes regret, $V(\bm{d}_{\gamma},p_{\Delta})$,
is the same as the frequentist regret at each support point, i.e.,
\begin{equation}
V(\bm{d}_{\gamma},p_{\Delta})=\frac{(\sigma_{1}+\sigma_{0})}{2}R(\gamma,\Delta).\label{eq:pf=000020Lemma=0000203=000020-=0000201}
\end{equation}
We aim to find a Nash equilibrium in a two-player game in which natures
chooses $p_{\Delta}$, equivalently $\Delta$, to maximize $R(\gamma,\Delta)$,
while the DM chooses $\bm{d}_{\gamma}$, equivalently $\gamma$, to
minimize $R(\gamma,\Delta)$. 

For $\eta=1$, the unique Nash equilibrium to this game is given by
$\Delta=\Delta_{0}^{*}$ and $\gamma=\gamma_{0}^{*}$. We start by
first demonstrating the existence of a Nash equilibrium. This is guaranteed
by Sion's minimax theorem \citep{sion1958general} as long as $R(\gamma,\Delta)$
is continuous in both arguments (which is easily verified), and `convex
quasi-concave' on $\mathbb{R}^{+}\times\mathbb{R}^{+}\backslash\{0\}$.\footnote{In fact, convexity can be replaced with quasi-convexity for the theorem.}
To show convexity in the first argument, write $R(\cdot,\Delta)=R_{1}(\alpha(\cdot,\Delta),\Delta)$
where 
\begin{align*}
R_{1}(\alpha,\Delta) & :=\Delta\alpha+\frac{2}{\Delta^{2}}(1-2\alpha)\ln\frac{1-\alpha}{\alpha};\ \textrm{and}\\
\alpha(\gamma,\Delta) & :=\frac{1-e^{-\Delta\gamma}}{e^{\Delta\gamma}-e^{-\Delta\gamma}}.
\end{align*}
Now, for any fixed $\Delta>0$, it is easy to verify that $R_{1}(\cdot,\Delta)$
and $\alpha(\cdot,\Delta)$ are convex over the domain $\mathbb{R}^{+}$.
Since the composition of convex functions is also convex, this proves
convexity of $R(\cdot,\Delta)$. To prove $R(\gamma,\cdot)$ is quasi-concave,
write $R(\gamma,\cdot)=R_{2}(\gamma,\alpha(\gamma,\cdot))$, where
\[
R_{2}(\gamma,\alpha):=\frac{1}{\gamma}\alpha\ln\frac{1-\alpha}{\alpha}+2\gamma^{2}\frac{(1-2\alpha)}{\ln\frac{1-\alpha}{\alpha}}.
\]
Now, $\alpha\ln\frac{1-\alpha}{\alpha}$ and $(1-2\alpha)/\ln\frac{1-\alpha}{\alpha}$
are concave over $\mathbb{R}^{+}\backslash\{0\}$, so $R_{2}(\gamma,\cdot)$
is also concave over $\mathbb{R}^{+}\backslash\{0\}$ for any fixed
$\gamma>0$. Concavity implies the level set $\left\{ \alpha:R_{2}(\gamma,\alpha)\ge\nu\right\} $
is a closed interval in $\mathbb{R}^{+}\backslash\{0\}$ for any $\nu\in\mathbb{R}$.
But $\alpha(\gamma,\cdot)$ is positive and strictly decreasing, so
for a fixed $\gamma>0$, 
\[
\left\{ \Delta:R(\gamma,\Delta)\ge\nu\right\} \equiv\left\{ \Delta:R_{2}(\gamma,\alpha(\gamma,\Delta))\ge\nu\right\} 
\]
is also a closed interval in $\mathbb{R}^{+}\backslash\{0\}$, and
therefore, convex, for any $\nu\in\mathbb{R}$. This proves quasi-concavity
of $R(\gamma,\cdot)$ whenever $\gamma>0$. At the same time, $R(\gamma,\Delta)=\Delta/2$
when $\gamma=0$; hence, $R(\gamma,\cdot)$ is in fact quasi-concave
for any $\gamma\ge0$. We thus conclude by Sion's theorem that a Nash
equilibrium exists. It is then routine to numerically compute $\Delta_{0}^{*},\gamma_{0}^{*}$
through first-order conditions and show that these values are unique;
we skip these calculations, which are straightforward. Figure \ref{fig:pf=000020Thm1}
provides a graphical illustration of the Nash equilibrium.

It remains to determine the Nash equilibrium under general $\eta$.
By the form of $R(\gamma,\Delta)$, if $\gamma_{0}^{*}$ is a best
response to $\Delta_{0}^{*}$ for $\eta=1$, then $\eta^{-1}\gamma_{0}^{*}$
is a best response to $\eta\Delta_{0}^{*}$ for general $\eta$. Similarly,
if $\Delta_{0}^{*}$ is a best response to $\gamma_{0}^{*}$ for $\eta=1$,
then $\eta\Delta_{0}^{*}$ is a best response to $\eta^{-1}\gamma_{0}^{*}$
for general $\eta$. This proves $\Delta^{*}:=\eta\Delta_{0}^{*}$
and $\gamma^{*}:=\eta^{-1}\gamma_{0}^{*}$ is a Nash equilibrium in
the general case.
\end{proof}
\begin{figure}
\includegraphics[scale=0.7,height=6cm]{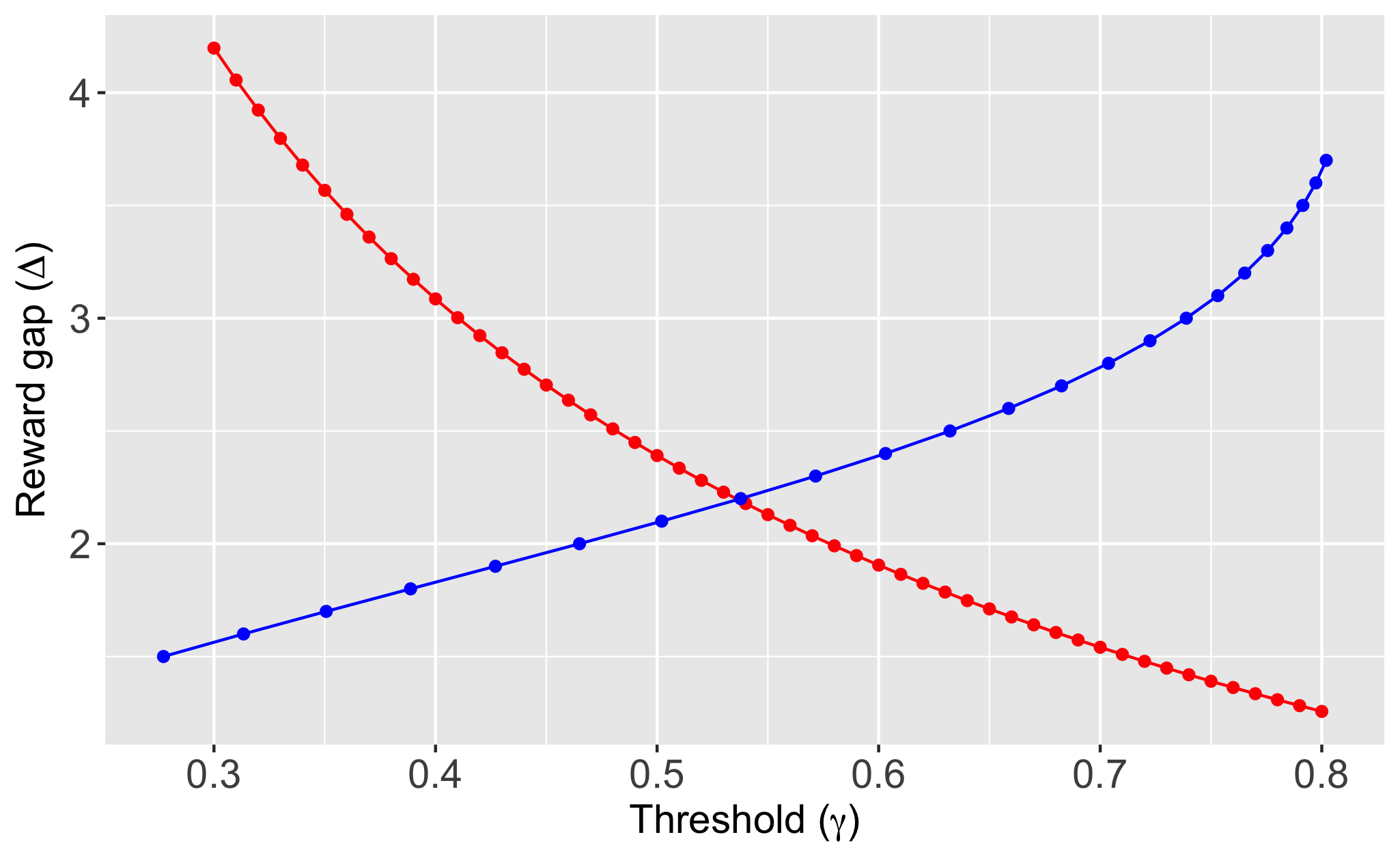}\\

\begin{raggedright}
{\scriptsize Note: The upward-sloping curve describes the best response
of $\Delta$ to a given $\gamma$, while the downward sloping curve
describes the best response of $\gamma$ to a given $\Delta$. The
point of intersection is the Nash equilibrium. This is for $\eta=1$.}{\scriptsize\par}
\par\end{raggedright}
\caption{Best responses and Nash equilibrium\protect\label{fig:pf=000020Thm1}}
\end{figure}

We now complete the proof of Theorem \ref{theorem=0000201}: By Lemmas
\ref{Lemma=0000201} and \ref{Lemma=0000203}, $\bm{d}^{*}$ is the
optimal Bayes decision corresponding to $p_{0}^{*}$. We now show
\begin{equation}
\sup_{\bm{\mu}}V(\bm{d}^{*},\bm{\mu})=V(\bm{d}^{*},p_{0}^{*}),\label{eq:pf=000020Theorem=0000201=000020-=000020verification=000020condition}
\end{equation}
which implies $\bm{d}^{*}$ is minimax optimal according to the verification
theorem in \citet[Theorem 17]{berger2013statistical}. To this end,
recall from Lemma \ref{Lemma=0000202} that the frequentist regret
$V(\bm{d}^{*},\bm{\mu})$ depends on $\bm{\mu}$ only through $\Delta:=2\vert\mu_{1}-\mu_{0}\vert/(\sigma_{1}+\sigma_{0})$.
Furthermore, by Lemma \ref{Lemma=0000203}, $\Delta^{*}$ is the best
response of nature to $\bm{d}^{*}$. These results imply
\[
\sup_{\bm{\mu}}V(\bm{d}^{*},\bm{\mu})=\frac{(\sigma_{1}+\sigma_{0})}{2}\sup_{\Delta}R(\gamma^{*},\Delta)=\frac{(\sigma_{1}+\sigma_{0})}{2}R(\gamma^{*},\Delta^{*}).
\]
But by (\ref{eq:pf=000020Lemma=0000203=000020-=0000201}), we also
have $V(\bm{d}^{*},p_{0}^{*})=\frac{(\sigma_{1}+\sigma_{0})}{2}R(\gamma^{*},\Delta^{*})$.
This proves (\ref{eq:pf=000020Theorem=0000201=000020-=000020verification=000020condition}). 

\subsection{Proof of Corollary \ref{Corollary=0000201}}

We employ the same strategy as in the proof of Theorem \ref{theorem=0000201}.
Suppose nature employs the indifference prior $p_{\Delta}$, for any
$\Delta>0$. Then by similar arguments as earlier, the DM is indifferent
between any sampling rule $\pi$, and the optimal implementation rule
is $\delta^{*}=\mathbb{I}\left\{ \frac{x_{1}(1)}{\sigma_{1}}-\frac{x_{0}(1)}{\sigma_{0}}\ge0\right\} .$ 

We now determine Nature's best response to the DM choosing $\bm{d}^{*}=(\pi^{*},\delta^{*})$,
where $\pi^{*}$ is the Neyman allocation. Consider an arbitrary $\bm{\mu}=(\mu_{1},\mu_{0})$
such that $\vert\mu_{1}-\mu_{0}\vert=\frac{\sigma_{1}+\sigma_{0}}{2}\Delta$.
Suppose $\mu_{1}>\mu_{0}$. Under $\pi^{*}$,
\[
\frac{dx_{1}(t)}{\sigma_{1}}-\frac{dx_{0}(t)}{\sigma_{0}}=\frac{\Delta}{2}dt+d\tilde{W}(t),
\]
where $\tilde{W}(\cdot)$ is the standard Wiener process, so the expected
regret under $\bm{d}^{*},\bm{\mu}$ is 
\begin{align}
V\left(\bm{d}^{*},\bm{\mu}\right) & =(\mu_{1}-\mu_{0})\mathbb{P}\left(\frac{x_{1}(1)}{\sigma_{1}}-\frac{x_{0}(1)}{\sigma_{0}}\le0\right)=\frac{\sigma_{1}+\sigma_{0}}{2}\Delta\Phi\left(-\frac{\Delta}{2}\right).\label{eq:nature=000020BR=0000201}
\end{align}
An analogous argument shows that the same expression holds when $\mu_{0}>\mu_{1}$
as well. Consequently, nature's optimal choice of $\bm{\mu}$ is to
set $\Delta$ to $\bar{\Delta}^{*}=2\arg\max_{\delta}\delta\Phi\left(-\delta\right)$,
but is otherwise indifferent between any $\bm{\mu}$ such that $\vert\mu_{1}-\mu_{0}\vert=\frac{\sigma_{1}+\sigma_{0}}{2}\bar{\Delta}^{*}$.
Thus, $p_{\bar{\Delta}^{*}}$ is a best response by nature to the
DM's choice of $\bm{d}^{*}=(\pi^{*},\delta^{*})$. 

We have thereby shown $p_{\bar{\Delta}^{*}},\bm{d}^{*}$ form a Nash
equilibrium. That $\bm{d}^{*}$ is minimax optimal then follows by
similar arguments as in the proof of Theorem \ref{theorem=0000201}. 

\subsection{Proof of Theorem \ref{Thm:=000020Parametric=000020families=000020lower=000020bound}\protect\label{subsec:Proof-of-Theorem}}

Our aim is to show (\ref{eq:lower=000020bound=000020-=000020intermediate=000020result}).
The outline of the proof is as follows: First, as in \citet{adusumilli2021risk},
we replace the true marginal and posterior distributions with suitable
approximations. Next, we apply dynamic programming arguments and viscosity
solution techniques to obtain a HJB-variational inequality (HJB-VI)
for the value function in the experiment. Finally, the HJB-VI is connected
to the problem of determining the optimal stopping time under diffusion
asymptotics. 

\subsubsection*{Step 0 (Definitions and preliminary observations)}

Under $m_{0}^{*}$, let $\gamma=1$ denote the event $\bm{h}=(h_{1}^{*},-h_{0}^{*})$
and $\gamma=0$ the event $\bm{h}=(-h_{1}^{*},h_{0}^{*})$. Also,
let ${\bf y}_{nq}^{(a)}:=\{Y_{i}^{(a)}\}_{i=1}^{\left\lfloor nq\right\rfloor }$
denote the stacked representation of outcomes $Y_{i}^{(a)}$ from
the first $nq$ observations corresponding to treatment $a$, and
for any $\bm{h}:=(h_{1},h_{0})$, take $P_{nq_{1},nq_{0},\bm{h}}$
to be the distribution corresponding to the joint density $p_{nq_{1},h_{1}}({\bf y}_{nq_{1}}^{(1)})\cdot p_{nq_{0},h_{1}}({\bf y}_{nq_{0}}^{(0)})$,
where 
\[
p_{nq,h_{a}}({\bf y}_{nq_{1}}^{(a)}):=\prod_{i=1}^{nq}p_{n,h_{a}}(Y_{i}^{(a)}).
\]
Also, define $\bar{P}_{n}$ as the marginal distribution of $\left({\bf y}_{nT}^{(1)},{\bf y}_{nT}^{(0)}\right)$,
i.e., it is the probability measure whose density, with respect to
the dominating measure $\nu({\bf y}_{nT}^{(1)},{\bf y}_{nT}^{(0)}):=\prod_{a\in\{0,1\}}\nu(Y_{1}^{(a)})\times\dots\times\nu(Y_{nT}^{(a)})$,
is
\[
\bar{p}_{n}\left({\bf y}_{nT}^{(1)},{\bf y}_{nT}^{(0)}\right)=\int p_{nT,h_{1}}({\bf y}_{nT}^{(1)})\cdot p_{nT,h_{1}}({\bf y}_{nT}^{(0)})dm_{0}^{*}(\bm{h}).
\]

Due to the two-point support of $m_{0}^{*}$, the posterior density
$p_{n}(\cdot\vert\xi_{t})$ can be associated with a scalar, 
\[
m_{n}(\xi_{t}):=m_{n}\left({\bf y}_{nq_{1}(t)}^{(1)},{\bf y}_{nq_{0}(t)}^{(0)}\right):=P_{n}\left(\gamma=1\vert{\bf y}_{nq_{1}(t)}^{(1)},{\bf y}_{nq_{0}(t)}^{(0)}\right).
\]
That the posterior depends on $\xi_{t}$ only via ${\bf y}_{nq_{1}(t)}^{(1)},{\bf y}_{nq_{0}(t)}^{(0)}$
is an immediate consequence of \citet[Lemma 1]{adusumilli2021risk}.
Recalling the definition of $\varpi_{n}(\cdot)$ in (\ref{eq:utility=000020function}),
we have $\varpi_{n}(\xi_{t})=\varpi_{n}(m_{n}(\xi_{t}))$, where,
for any $m\in[0,1]$,
\begin{align*}
\varpi_{n}(m) & :=\min\left\{ \left\{ \mu_{n,0}(h_{0}^{*})-\mu_{n,1}(-h_{1}^{*})\right\} (1-m),\left\{ \mu_{n,1}(h_{1}^{*})-\mu_{n,0}(-h_{0}^{*})\right\} m\right\} \\
 & =\left(\mu_{n,1}(h_{1}^{*})-\mu_{n,0}(-h_{0}^{*})\right)\min\{m,1-m\}.
\end{align*}
The first equation above always holds, while the second holds under
the simplification $\mu_{n,a}(h)=-\mu_{n,a}(-h)$ described in Section
\ref{sec:Formal-properties}. 

Let 
\begin{equation}
z_{a,nq_{a}}:=\frac{I_{a}^{-1/2}}{\sqrt{n}}\sum_{i=1}^{\left\lfloor nq_{a}\right\rfloor }\psi_{a}(Y_{i}^{(a)}),\label{eq:score=000020process=000020definition}
\end{equation}
denote the (standardized) score process. Under quadratic mean differentiability
- Assumption 1(i) - the following SLAN property holds for both treatments:
\begin{equation}
\sum_{i=1}^{\left\lfloor nq_{a}\right\rfloor }\ln\frac{dp_{\theta_{0}+h/\sqrt{n}}^{(a)}}{dp_{\theta_{0}}^{(a)}}(Y_{i}^{(a)})=h^{\intercal}I_{a}^{1/2}z_{a,nq_{a}}-\frac{q_{a}}{2}h^{\intercal}I_{a}h+o_{P_{nT,\theta_{0}}^{(a)}}(1),\ \textrm{uniformly over bounded }q_{a}.\label{eq:LAN}
\end{equation}
See \citet[Lemma 2]{adusumilli2021risk} for the proof.\footnote{It should be noted that the score process in that paper is defined
slightly differently, as $I_{a}^{-1/2}z_{a,nq_{a}}$ under the present
notation.} 

As in \citet{adusumilli2021risk}, we now define approximate versions
of the true marginal and posterior by replacing the actual likelihood
$\prod_{a}p_{nq_{a},h_{a}}^{(a)}({\bf y}_{nT}^{(a)})$ with
\begin{align}
\prod_{a}\lambda_{nq,h_{a}}^{(a)}({\bf y}_{nq}^{(a)}) & :=\prod_{a}\frac{d\Lambda_{nq,h_{a}}^{(a)}({\bf y}_{nq}^{(a)})}{d\nu},\ \textrm{where }\nonumber \\
\lambda_{nq,h}^{(a)}({\bf y}_{nq}^{(a)}) & :=\exp\left\{ h^{\intercal}I_{a}^{1/2}z_{a,nq}-\frac{q}{2}h^{\intercal}I_{a}h\right\} p_{nq,\theta_{0}}^{(a)}({\bf y}_{nq}^{(a)})\ \forall\ q\in[0,T].\label{eq:tilted=000020measure}
\end{align}
In other words, we approximate the true likelihood with the first
two terms in the SLAN expansion (\ref{eq:LAN}). The construction
of the approximate marginal and posterior is described below. 

\textit{(Approximate marginal:)} Denote by $\tilde{P}_{nq_{1},nq_{0},\bm{h}}$
the measure whose density is $\lambda_{nq_{1},h_{1}}^{(1)}({\bf y}_{nq_{1}}^{(1)})\cdot\lambda_{nq_{0},h_{0}}^{(0)}({\bf y}_{nq_{0}}^{(0)})$,
and take $\tilde{\bar{P}}_{nq_{1},nq_{0}}$ to be its marginal over
${\bf y}_{nq_{1}}^{(1)},{\bf y}_{nq_{0}}^{(0)}$ given the prior $m_{0}^{*}(\bm{h})$.
Note that the density (wrt $\nu$) of $\tilde{\bar{P}}_{nq_{1},nq_{0}}$
is 
\begin{equation}
\tilde{\bar{p}}_{nq_{1},nq_{0}}\left({\bf y}_{nq_{1}}^{(1)},{\bf y}_{nq_{0}}^{(0)}\right)=\int\lambda_{nq_{1},h^{(1)}}^{(1)}\left({\bf y}_{nq_{1}}^{(1)}\right)\cdot\lambda_{nq_{0},h^{(0)}}^{(0)}\left({\bf y}_{nq_{0}}^{(0)}\right)dm_{0}^{*}(\bm{h}).\label{eq:marginal=000020density=000020definition}
\end{equation}
Also, define $\tilde{\bar{p}}_{n}\left({\bf y}_{nT}^{(1)},{\bf y}_{nT}^{(0)}\right):=\tilde{\bar{p}}_{nT,nT}\left({\bf y}_{nT}^{(1)},{\bf y}_{nT}^{(0)}\right)$.
Then, $\tilde{\bar{p}}_{n}\left({\bf y}_{nT}^{(1)},{\bf y}_{nT}^{(0)}\right)$
approximates the true marginal $\bar{p}_{n}\left({\bf y}_{nT}^{(1)},{\bf y}_{nT}^{(0)}\right)$.

\textit{(Approximate posterior:) }Next, let $\tilde{\varphi}(t)$
be the approximate likelihood ratio
\[
\tilde{\varphi}(t)=\frac{\lambda_{nq_{1},h_{1}^{*}}^{(1)}\left({\bf y}_{nq_{1}(t)}^{(1)}\right)\cdot\lambda_{nq_{0},-h_{0}^{*}}^{(0)}\left({\bf y}_{nq_{0}(t)}^{(0)}\right)}{\lambda_{nq_{1},-h_{1}^{*}}^{(1)}\left({\bf y}_{nq_{1}(t)}^{(1)}\right)\cdot\lambda_{nq_{0},h_{0}^{*}}^{(0)}\left({\bf y}_{nq_{0}(t)}^{(0)}\right)}=\exp\left\{ \Delta^{*}\rho_{n}(t)\right\} ,
\]
where
\begin{equation}
\rho_{n}(t):=\frac{\dot{\mu}_{1}^{\intercal}z_{1,nq_{1}(t)}}{\sigma_{1}}-\frac{\dot{\mu}_{0}^{\intercal}z_{0,nq_{0}(t)}}{\sigma_{0}}.\label{eq:Definition=000020of=000020rho}
\end{equation}
Based on the above, we can approximate the true posterior, $m_{n}(\xi_{t})$,
by
\begin{align}
\frac{\tilde{\varphi}(t)}{1+\tilde{\varphi}(t)}=\frac{\exp\left\{ \Delta^{*}\rho_{n}(t)\right\} }{1+\exp\left\{ \Delta^{*}\rho_{n}(t)\right\} }:=\tilde{m}(\rho_{n}(t)),\label{eq:approximate=000020posterior}
\end{align}
where $\tilde{m}(\rho):=\exp(\Delta^{*}\rho)/(1+\exp(\Delta^{*}\rho))$
for $\rho\in\mathbb{R}$. When $\rho_{n}(t)=\rho$, the approximate
posterior $\tilde{m}(\rho)$ in turn implies an approximate posterior,
$\tilde{p}_{n}(\bm{h}\vert\rho)$, over $\bm{h}$ that takes the value
$(h_{1}^{*},-h_{0}^{*})$ with probability $\tilde{m}(\rho)$ and
$(-h_{1}^{*},h_{0}^{*})$ with probability $1-\tilde{m}(\rho)$. 

\subsubsection*{Step 1 (Posterior and probability approximations)}

Set $V_{n,T}^{*}=\inf_{\bm{d}\in\mathcal{D}_{n,T}}V_{n}^{*}(\bm{d},m_{0}^{*})$.
Using dynamic programming arguments, it is straightforward to show
that there exists a non-randomized sampling rule and stopping time
that minimizes $V_{n}^{*}(\bm{d},m_{0})$ for any prior $m_{0}$.
We therefore restrict $\mathcal{D}_{n,T}$ to the set of all deterministic
rules, $\mathcal{\bar{D}}_{n,T}$. Under deterministic policies, the
sampling rules $\pi_{nt}$, states $\xi_{t}$ and stopping times $\tau$
are all deterministic functions of ${\bf y}_{nT}^{(1)},{\bf y}_{nT}^{(0)}$.
Recall that ${\bf y}_{nT}^{(1)},{\bf y}_{nT}^{(0)}$ are the stacked
vector of outcomes under $nT$ observations of each treatment. It
is useful to think of $\{\pi_{nt}\}_{t=1/n}^{T},\tau$ as quantities
mapping $({\bf y}_{nT}^{(1)},{\bf y}_{nT}^{(0)})$ to realizations
of regret.\footnote{Note that $\pi,\tau$ still need to satisfy the measurability restrictions,
and some components of ${\bf y}_{nT}^{(a)}$ may not be observed as
both treatments cannot be sampled $nT$ times.} Taking $\bar{\mathbb{E}}_{n}[\cdot]$ to be the expectation under
$\bar{P}_{n}$, we then have
\[
V_{n}^{*}(\bm{d},m_{0}^{*})=\mathbb{\bar{E}}_{n}\left[\sqrt{n}\varpi_{n}\left(m_{n}\left(\xi_{\tau}\right)\right)+c\tau\right],
\]
for any deterministic $\bm{d}\in\mathcal{\bar{D}}_{n,T}$. 

Now, take $\tilde{\bar{\mathbb{E}}}_{n}[\cdot]$ to be the expectation
under $\tilde{\bar{P}}_{n}$, and define 
\begin{equation}
\tilde{V}_{n}(\bm{d},m_{0}^{*})=\mathbb{\tilde{\bar{E}}}_{n}\left[\sqrt{n}\varpi_{n}\left(\tilde{m}\left(\rho_{n}(\tau)\right)\right)+c\tau\right].\label{eq:definition=000020of=000020tilde=000020V}
\end{equation}
By Lemma 7 in Appendix F,
\[
\lim_{n\to\infty}\sup_{\bm{d}\in\bar{\mathcal{D}}_{n,T}}\left|V_{n}^{*}(\bm{d},m_{0}^{*})-\tilde{V}_{n}(\bm{d},m_{0}^{*})\right|=0.
\]
This in turn implies $\lim_{n\to\infty}\left|V_{n,T}^{*}-\tilde{V}_{n,T}^{*}\right|=0$,
where $\tilde{V}_{n,T}^{*}:=\inf_{\bm{d}\in\mathcal{\bar{D}}_{n,T}}\tilde{V}_{n}^{*}(\bm{d},m_{0}^{*})$. 

\subsubsection*{Step 2 (Recursive formula for $\tilde{V}_{n,T}^{*}$)}

We now employ dynamic programming arguments to obtain a recursion
for $\tilde{V}_{n,T}^{*}$. This requires a bit of care since $\tilde{\bar{P}}_{n}$
is not a probability, even though it does integrate to 1 asymptotically. 

Recall that $\tilde{p}_{n}(\bm{h}\vert\rho)$ is the probability measure
on $\bm{h}$ that assigns probability $\tilde{m}(\rho)$ to $(h_{1}^{*},-h_{0}^{*})$
and probability $1-\tilde{m}(\rho)$ to $(-h_{1}^{*},h_{0}^{*})$.
Define 
\begin{align}
\tilde{p}_{n}(Y^{(a)}\vert\rho) & =p_{\theta_{0}}^{(a)}(Y^{(a)})\cdot\int\exp\left\{ \frac{1}{\sqrt{n}}h_{a}^{\intercal}\psi_{a}(Y^{(a)})-\frac{1}{2n}h_{a}^{\intercal}I_{a}h_{a}\right\} d\tilde{p}_{n}(\bm{h}\vert\rho),\nonumber \\
\tilde{\bar{p}}_{n}({\bf y}_{-nq_{1}}^{(1)},{\bf y}_{-nq_{0}}^{(0)}\vert\rho,q_{1},q_{0}) & =\int\frac{\lambda_{nT,h_{1}}^{(1)}\left({\bf y}_{nT}^{(1)}\right)\cdot\lambda_{nT,h_{0}}^{(0)}\left({\bf y}_{nT}^{(0)}\right)}{\lambda_{nq_{1},h_{1}}^{(1)}\left({\bf y}_{nq_{1}}^{(1)}\right)\cdot\lambda_{nq_{0},h_{0}}^{(0)}\left({\bf y}_{nq_{0}}^{(0)}\right)}d\tilde{p}_{n}(\bm{h}\vert\rho),\quad\textrm{and}\nonumber \\
\eta(\rho,q_{1},q_{0}) & =\int d\tilde{\bar{p}}_{n}\left({\bf y}_{-nq_{1}}^{(1)},{\bf y}_{-nq_{0}}^{(0)}\vert\rho,q_{1},q_{0}\right),\label{eq:definitions}
\end{align}
where ${\bf y}_{-nq}^{(a)}:=\{Y_{nq+1}^{(a)},\dots,Y_{nT}^{(a)}\}$.
In words, $\tilde{\bar{p}}_{n}({\bf y}_{-nq_{1}}^{(1)},{\bf y}_{-nq_{0}}^{(0)}\vert\rho,q_{1},q_{0})$
is the approximate probability density over the future values of the
stacked rewards $\{Y_{i}^{(a)}\}_{i=nq_{a}+1}^{nT}$ given the current
state $\rho,q_{1},q_{0}$. Note that $\eta(\rho,q_{1},q_{0})$ is
the normalization constant of $\tilde{\bar{p}}_{n}({\bf y}_{-nq_{1}}^{(1)},{\bf y}_{-nq_{0}}^{(0)}\vert\rho,q_{1},q_{0})$.

By Lemma 8 in Appendix F, $\tilde{V}_{n,T}^{*}=\tilde{V}_{n,T}^{*}(0,0,0,0)$,
where $\tilde{V}_{n,T}^{*}(\cdot)$ solves the recursion
\begin{align}
 & \tilde{V}_{n,T}^{*}\left(\rho,q_{1},q_{0},t\right)=\min\bigg\{\sqrt{n}\eta(\rho,q_{1},q_{0})\varpi_{n}(\tilde{m}(\rho)),\nonumber \\
 & \left.\frac{\eta(\rho,q_{1},q_{0})c}{n}+\min_{a\in\{0,1\}}\int\tilde{V}_{n,T}^{*}\left(\rho+\frac{(2a-1)\dot{\mu}_{a}^{\intercal}I_{a}^{-1}\psi_{a}(Y^{(a)})}{\sqrt{n}\sigma_{a}},q_{1}+\frac{a}{n},q_{0}+\frac{1-a}{n},t+\frac{1}{n}\right)d\tilde{p}_{n}(Y^{(a)}\vert\rho)\right\} ,\label{eq:recursion=000020for=000020tilde=000020V}
\end{align}
for $t\le T$, and 
\[
\tilde{V}_{n,T}^{*}\left(\rho,q_{1},q_{0},T\right)=\sqrt{n}\eta(\rho,q_{1},q_{0})\varpi_{n}(\tilde{m}(\rho)).
\]
The function $\eta(\cdot)$ accounts for the fact $\tilde{\bar{P}}_{n}$
is not a probability. 

Now, Lemma 9 in Appendix F shows that 
\begin{equation}
\sup_{\rho,q_{1},q_{0}}\left|\eta(\rho,q_{1},q_{0})-1\right|\le Mn^{-\vartheta}\label{eq:normalization=000020constant=000020bound}
\end{equation}
for some $M<\infty$ and any $\vartheta\in(0,1/2)$. Furthermore,
by Assumption 1(iii), 
\begin{equation}
\lim_{n\to\infty}\sup_{m\in[0,1]}\left|\sqrt{n}\varpi_{n}(m)-\varpi(m)\right|=0,\label{eq:pf:=000020Thm=0000202:=000020Bayes=000020risk=000020at=000020conclusion=000020of=000020experiment}
\end{equation}
where $\varpi(m):=\frac{\sigma_{1}+\sigma_{0}}{2}\Delta^{*}\min\{m,1-m\}$.
Since $\varpi(\cdot)$ is uniformly bounded, it follows from (\ref{eq:pf:=000020Thm=0000202:=000020Bayes=000020risk=000020at=000020conclusion=000020of=000020experiment})
that $\sqrt{n}\varpi_{n}(\cdot)$ is also uniformly bounded. Then,
(\ref{eq:normalization=000020constant=000020bound}) and (\ref{eq:pf:=000020Thm=0000202:=000020Bayes=000020risk=000020at=000020conclusion=000020of=000020experiment})
imply
\[
\lim_{n\to\infty}\left|\tilde{V}_{n,T}^{*}(0)-\breve{V}_{n,T}^{*}(0)\right|=0,
\]
where $\breve{V}_{n,T}(\rho,t)$ is defined as the solution to the
recursion 
\begin{align}
\breve{V}_{n,T}^{*}\left(\rho,t\right) & =\min\left\{ \varpi(\tilde{m}(\rho)),\frac{c}{n}+\min_{a\in\{0,1\}}\int\breve{V}_{n,T}^{*}\left(\rho+\frac{(2a-1)\dot{\mu}_{a}^{\intercal}I_{a}^{-1}\psi_{a}(Y^{(a)})}{\sqrt{n}\sigma_{a}},t+\frac{1}{n}\right)d\tilde{p}_{n}(Y^{(a)}\vert\rho)\right\} \nonumber \\
 & \quad\textrm{for }t\le T,\label{eq:recursive=000020formulation}\\
\breve{V}_{n,T}^{*}\left(\rho,T\right) & =\varpi(\tilde{m}(\rho)).\nonumber 
\end{align}
We can drop the state variables $q_{1},q_{0}$ in $\breve{V}_{n,T}^{*}\left(\cdot\right)$
as they enter the definition of $\tilde{V}_{n,T}^{*}\left(\rho,q_{1},q_{0},t\right)$
only via $\eta(\rho,q_{1},q_{0})$, which was shown in (\ref{eq:normalization=000020constant=000020bound})
to be uniformly close to 1. 

\subsubsection*{Step 3 (PDE approximation and relationship to optimal stopping)}

For any $\rho\in\mathbb{R}$, let
\[
\varpi(\rho):=\varpi(\tilde{m}(\rho))=\frac{(\sigma_{1}+\sigma_{0})\Delta^{*}}{2}\min\left\{ \frac{\exp(\Delta^{*}\rho)}{1+\exp(\Delta^{*}\rho)},\frac{1}{1+\exp(\Delta^{*}\rho)}\right\} .
\]
Lemma 10 in Appendix F shows that $\breve{V}_{n,T}^{*}(\cdot)$ converges
locally uniformly to $V_{T}^{*}(\cdot)$, the unique viscosity solution
of the HJB-VI 
\begin{align}
\min\left\{ \varpi(\rho)-V_{T}^{*}(\rho,t),c+\partial_{t}V_{T}^{*}+\frac{\Delta^{*}}{2}(2\tilde{m}(\rho)-1)\partial_{\rho}V_{T}^{*}+\frac{1}{2}\partial_{\rho}^{2}V_{T}^{*}\right\}  & =0\ \textrm{for }t\le T,\nonumber \\
V_{T}^{*}(\rho,T) & =\varpi(\rho).\label{eq:HJB-VI-pf:Thm2}
\end{align}
Note that the sampling rule does not enter the HJB-VI. This is a consequence
of the choice of the prior, $m_{0}^{*}$. 

There is a well known connection between HJB-VIs and the problem of
optimal stopping that goes by the name of smooth-pasting or the high
contact principle, see \citet[Chapter 10]{oksendal2003stochastic}
for an overview. In the present context, letting $W(t)$ denote one-dimensional
Brownian motion, it follows by \citet{reikvam1998viscosity} that
\begin{align*}
V_{T}^{*}(0,0) & =\inf_{\tau\le T}\mathbb{E}\left[\varpi(\rho_{\tau})+c\tau\right],\ \textrm{where}\\
d\rho_{t} & =\frac{\Delta^{*}}{2}(2\tilde{m}(\rho_{t})-1)dt+dW(t);\ \rho_{0}=0,
\end{align*}
and $\tau$ is the set of all stopping times adapted to the filtration
$\mathcal{F}_{t}$ generated by $\rho_{t}$. 

\subsubsection*{Step 4 (Taking $T\to\infty$)}

Through steps 1-3, we have shown 
\[
\lim_{n\to\infty}\inf_{\bm{d}\in\mathcal{D}_{n,T}}\sup_{\bm{h}}V_{n}(\bm{d},\bm{h})\ge\lim_{n\to\infty}\inf_{\bm{d}\in\mathcal{D}_{n,T}}V_{n}(\bm{d},m_{0}^{*})=V_{T}^{*}(0,0).
\]

We now argue that 
\[
\lim_{T\to\infty}V_{T}^{*}(0,0)=V_{\infty}^{*}:=\inf_{\tau}\mathbb{E}\left[\varpi(\rho_{\tau})+c\tau\right].
\]
Suppose not: Then, there exists $\epsilon>0$, a sequence $\{T_{j}\}_{j}$
with $T_{j}\uparrow\infty$, and some stopping time $\bar{\tau}$
such that $V(\bar{\tau}):=\mathbb{E}\left[\varpi(\rho_{\bar{\tau}})+c\bar{\tau}\right]<V_{T_{j}}^{*}(0,0)-\epsilon$
for all $j$ (note that we always have $V_{T}^{*}(0,0)\ge V_{\infty}^{*}$
by definition). Now, $\varpi(\cdot)$ is uniformly bounded, so by
the dominated convergence theorem, $\lim_{j\to\infty}\mathbb{E}\left[\varpi(\rho_{\bar{\tau}\wedge T_{j}})\right]=\mathbb{E}\left[\varpi(\rho_{\bar{\tau}})\right]$.
Hence, 
\begin{align*}
\lim_{j\to\infty}V_{T_{j}}^{*}(0,0) & \le\lim_{j\to\infty}\mathbb{E}\left[\varpi(\rho_{\bar{\tau}\wedge T_{j}})+c\left(\bar{\tau}\wedge T_{j}\right)\right]\\
 & =\mathbb{E}\left[\varpi(\rho_{\bar{\tau}})\right]+\lim_{j\to\infty}c\mathbb{E}\left[\bar{\tau}\wedge T_{j}\right]\le V(\bar{\tau}).
\end{align*}
This is a contradiction. 

It remains to show $V_{\infty}^{*}$ is the same as $V^{*}$, the
value of the two-player game in Theorem \ref{theorem=0000201}. Define
\[
m_{t}=\frac{\exp(\Delta^{*}\rho_{t})}{1+\exp(\Delta^{*}\rho_{t})}.
\]
By a change of variables from $\rho_{t}$ to $m_{t}$, we can write
$V_{\infty}^{*}:=\inf_{\tau}\mathbb{E}\left[\varpi(m_{t})+c\tau\right]$,
where $dm_{t}=\Delta^{*}m_{t}(1-m_{t})dW_{t}$ by Ito's lemma. But
by way of the proof of Lemma \ref{Lemma=0000201}, see (\ref{eq:pf=000020thm=0000201:=000020objective=000020fn=000020for=000020stopping=000020rule}),
this is just $V^{*}$. The theorem can therefore be considered proved. 

\subsection{Proof of Theorem \ref{Thm:=000020Parametric=000020-=000020attaining=000020the=000020bound}}

For any $\bm{h}=(h_{1},h_{0})$, let $P_{n,\bm{h}}$ denote the joint
distribution with density $p_{nT,\theta_{0}+h_{1}/\sqrt{n}}^{(1)}({\bf y}_{nT}^{(1)})\cdot p_{nT,\theta_{0}+h_{0}/\sqrt{n}}^{(0)}({\bf y}_{nT}^{(0)})$.
Take $\mathbb{E}_{n,\bm{h}}[\cdot]$ to be the corresponding expectation.
We can write $V_{n}(\bm{d}_{n,T},\bm{h})$ as 
\[
V_{n}(\bm{d}_{n,T},\bm{h})=\mathbb{E}_{n,\bm{h}}\left[\sqrt{n}\left(\mu_{n,1}(h_{1})-\mu_{n,0}(h_{0})\right)\mathbb{I}\{\delta_{n,T}\ge0\}+c\tau_{n,T}\right].
\]
Define $\mu(\bm{h})=\left(\dot{\mu}_{1}^{\intercal}h_{1},\dot{\mu}_{0}^{\intercal}h_{0}\right)$,
$\Delta\mu(\bm{h})=\dot{\mu}_{1}^{\intercal}h_{1}-\dot{\mu}_{0}^{\intercal}h_{0}$
and $\Delta_{n}\mu(\bm{h})=\mu_{n,1}(h_{1})-\mu_{n,0}(h_{0})$. In
addition, we also define $\tilde{q}_{a}(t):=\sigma_{a}t/(\sigma_{1}+\sigma_{0})$.

\subsubsection*{Step 1 (Weak convergence of $\rho_{n}(t)$)}

Denote $P_{n,0}=P_{n,(0,0)}.$ By the SLAN property (\ref{eq:LAN}),
independence of ${\bf y}_{nT}^{(1)},{\bf y}_{n,T}^{(0)}$ given $\bm{h}$,
and the central limit theorem,
\begin{align}
\ln\frac{dP_{n,\bm{h}}}{dP_{n,0}}\left({\bf y}_{nT}^{(1)},{\bf y}_{n,T}^{(0)}\right) & =\sum_{a\in\{0,1\}}\left\{ h_{a}^{\intercal}I_{a}^{1/2}z_{a,nT}-\frac{T}{2}h_{a}^{\intercal}I_{a}h_{a}\right\} +o_{Pn,0}(1)\label{eq:AR-process=000020-0}\\
 & \xrightarrow[P_{n,0}]{d}\mathcal{N}\left(\frac{-T}{2}\sum_{a\in\{0,1\}}h_{a}^{\intercal}I_{a}h_{a},T\sum_{a\in\{0,1\}}h_{a}^{\intercal}I_{a}h_{a}\right).\label{eq:LR=000020process-1}
\end{align}
Therefore, by Le Cam's first lemma, $P_{n,\bm{h}}$ and $P_{n,0}$
are mutually contiguous. 

We now determine the distribution of $\rho_{n}(t)$. We start by showing
\begin{equation}
\left|\frac{\dot{\mu}_{a}^{\intercal}I_{a}^{-1}}{\sigma_{a}\sqrt{n}}\sum_{i=1}^{\left\lfloor nq_{a}(t)\right\rfloor }\psi_{a}(Y_{i}^{(a)})-\frac{\dot{\mu}_{a}^{\intercal}I_{a}^{-1}}{\sigma_{a}\sqrt{n}}\sum_{i=1}^{\left\lfloor n\tilde{q}_{a}(t)\right\rfloor }\psi_{a}(Y_{i}^{(a)})\right|=o_{P_{n,0}}(1),\label{eq:approximation=000020of=000020rho_n:=000020step=0000201}
\end{equation}
uniformly over $t\le T$. Choose any $b\in(1/2,1)$. For $t\le n^{-b}$,
we must have $q_{a}(t),\tilde{q}_{a}(t)\le n^{-b}$, so (\ref{eq:approximation=000020of=000020rho_n:=000020step=0000201})
follows from Assumption 1(ii), which implies
\begin{equation}
\sup_{1\le i\le nT}\vert\psi_{a}(Y_{i}^{(a)})\vert=O_{P_{n,0}}(n^{1/r}),\ \textrm{for any }r>0.\label{eq:pf:proof=000020Theorem=0000203}
\end{equation}
As for the other values of $t$, by (\ref{eq:requirement=000020on=000020pi_n})
and (\ref{eq:pf:proof=000020Theorem=0000203}),
\begin{align*}
\frac{\dot{\mu}_{a}^{\intercal}I_{a}^{-1}}{\sigma_{a}\sqrt{n}}\left\{ \sum_{i=1}^{\left\lfloor nq_{a}(t)\right\rfloor }\psi_{a}(Y_{i}^{(a)})-\sum_{i=1}^{\left\lfloor n\tilde{q}_{a}(t)\right\rfloor }\psi_{a}(Y_{i}^{(a)})\right\}  & \lesssim\sqrt{n}\left|q_{a}(t)-\tilde{q}_{a}(t)\right|\sup_{1\le i\le nT}\vert\psi_{a}(Y_{i}^{(a)})\vert=o_{P_{n,0}}(1),
\end{align*}
uniformly over $t\in(n^{-b},T]$. 

Now, (\ref{eq:approximation=000020of=000020rho_n:=000020step=0000201})
implies
\begin{equation}
\rho_{n}(t)=\frac{\dot{\mu}_{1}^{\intercal}I_{1}^{-1}}{\sigma_{1}\sqrt{n}}\sum_{i=1}^{\left\lfloor n\tilde{q}_{1}(t)\right\rfloor }\psi_{1}(Y_{i}^{(1)})-\frac{\dot{\mu}_{0}^{\intercal}I_{0}^{-1}}{\sigma_{0}\sqrt{n}}\sum_{i=1}^{\left\lfloor n\tilde{q}_{0}(t)\right\rfloor }\psi_{0}(Y_{i}^{(0)})+o_{P_{n,0}}(1)\ \textrm{uniformly over }t\le T.\label{eq:=000020approximation=000020of=000020rho_n=000020step=0000202}
\end{equation}
By Donsker's theorem, and recalling that $\tilde{q}_{a}(t)=\sigma_{a}t/(\sigma_{1}+\sigma_{0})$,
\[
\frac{\dot{\mu}_{a}^{\intercal}I_{a}^{-1}}{\sigma_{a}\sqrt{n}}\sum_{i=1}^{\left\lfloor n\tilde{q}_{a}(\cdot)\right\rfloor }\psi_{a}(Y_{i}^{(a)})\xrightarrow[P_{n,0}]{d}\sqrt{\frac{\sigma_{a}}{\sigma_{1}+\sigma_{0}}}W_{a}(\cdot),
\]
where $W_{1}(\cdot),W_{0}(\cdot)$ can be taken to be independent
Wiener processes due to the independence of ${\bf y}_{nT}^{(1)},{\bf y}_{n,T}^{(0)}$
under $P_{n,0}$. Combined with (\ref{eq:=000020approximation=000020of=000020rho_n=000020step=0000202}),
we conclude
\begin{equation}
\rho_{n}(\cdot)\xrightarrow[P_{n,0}]{d}\tilde{W}(\cdot),\label{eq:convergence=000020of=000020rho_n=000020under=000020P_0}
\end{equation}
where $\tilde{W}(\cdot)=\sqrt{\frac{\sigma_{1}}{\sigma_{1}+\sigma_{0}}}W_{1}(\cdot)-\sqrt{\frac{\sigma_{0}}{\sigma_{1}+\sigma_{0}}}W_{0}(\cdot)$
is another Wiener process. 

Let $Z$ denote the normal random variable in (\ref{eq:LR=000020process-1}).
Equations (\ref{eq:LR=000020process-1}) and (\ref{eq:convergence=000020of=000020rho_n=000020under=000020P_0})
imply that $\rho_{n}(\cdot),\ln\left(dP_{n,\bm{h}}/dP_{n,0}\right)$
are asymptotically tight, and therefore, the joint $\left(\rho_{n}(\cdot),\ln\left(dP_{n,\bm{h}}/dP_{n,0}\right)\right)$
is also asymptotically tight under $P_{n,0}.$ Furthermore, for any
$t\in[0,T]$, it can be shown using (\ref{eq:=000020approximation=000020of=000020rho_n=000020step=0000202})
and (\ref{eq:AR-process=000020-0}) that 
\[
\left(\begin{array}{c}
\rho_{n}(t)\\
\ln\frac{dP_{n,\bm{h}}}{dP_{n,0}}
\end{array}\right)\xrightarrow[P_{n,0}]{d}\left(\begin{array}{c}
\tilde{W}(t)\\
Z
\end{array}\right)\sim\mathcal{N}\left(\left(\begin{array}{c}
0\\
\frac{-T}{2}\sum_{a}h_{a}^{\intercal}I_{a}h_{a}
\end{array}\right),\left[\begin{array}{cc}
t & \frac{\Delta\mu(\bm{h})}{\sigma_{1}+\sigma_{0}}t\\
\frac{\Delta\mu(\bm{h})}{\sigma_{1}+\sigma_{0}}t & T\sum_{a}h_{a}^{\intercal}I_{a}h_{a}
\end{array}\right]\right).
\]
Based on the above, an application of Le Cam's third lemma as in \citet[Theorem 3.10.12]{van1996weak}
then gives 
\begin{equation}
\rho_{n}(\cdot)\xrightarrow[P_{n,\bm{h}}]{d}\rho(\cdot)\quad\textrm{where }\ \rho(t):=\frac{\Delta\mu(\bm{h})}{\sigma_{1}+\sigma_{0}}t+\tilde{W}(t).\label{eq:convergence=000020of=000020rho_n=000020general}
\end{equation}

\subsubsection*{Step 2 (Weak convergence of $\delta_{n,T},\tau_{n,T}$)}

Let $\mathbb{D}[0,T]$ denote the metric space of all functions from
$[0,T]$ to $\mathbb{R}$ equipped with the sup norm. For any element
$z(\cdot)\in\mathbb{D}[0,T]$, define $\tau_{T}(z)=T\wedge\inf\{t:\vert z(t)\vert\ge\gamma\}$
and $\delta_{T}(z)=\mathbb{I}\{z(\tau_{T}(z))>0\}$. 

Now, under $\bm{h}=(0,0)$, $\rho(\cdot)$ is the Wiener process,
whose sample paths take values (with probability 1) in $\bar{\mathbb{C}}[0,T]$,
the set of all continuous functions such that $\gamma,-\gamma$ are
regular points (i.e., if $z(t)=\gamma$, $z(\cdot)-\gamma$ changes
sign infinitely often in any time interval $[t,t+\epsilon]$, $\epsilon>0$;
a similar property holds under $z(t)=-\gamma$). The latter is a well
known property of Brownian motion, see \citet[Problem 2.7.18]{karatzas2012brownian},
and it implies $z(\cdot)\in\mathbb{\bar{C}}[0,T]$ must `cross' the
boundary within an arbitrarily small time interval after hitting $\gamma$
or $-\gamma$. It is then easy to verify that if $z_{n}\to z$ with
$z_{n}\in\mathbb{D}[0,T]$ for all $n$ and $z\in\mathbb{\bar{C}}[0,T]$,
then $\tau_{T}(z_{n})\to\tau_{T}(z)$ and $\delta_{T}(z_{n})\to\delta_{T}(z)$.
By construction, $\tau_{n,T}=\tau_{T}(\rho_{n})$ and $\delta_{n,T}=\delta_{T}(\rho_{n})$,
so by (\ref{eq:convergence=000020of=000020rho_n=000020under=000020P_0})
and the extended continuous mapping theorem \citep[Theorem 1.11.1]{van1996weak}
\[
(\tau_{n,T},\delta_{n,T})\xrightarrow[P_{n,0}]{d}(\tau_{T}^{*},\delta_{T}^{*}),
\]
where $\tau_{T}^{*}:=\tau_{T}(\rho)$ and $\delta_{T}^{*}:=\delta_{T}(\rho)$. 

For general $\bm{h}$, $\rho(\cdot)$ is distributed as in (\ref{eq:convergence=000020of=000020rho_n=000020general}).
By the Girsanov theorem, the probability law induced on $\mathbb{D}[0,T]$
by the process $\frac{\Delta\mu(\bm{h})}{\sigma_{1}+\sigma_{0}}t+\tilde{W}(t)$
is absolutely continuous with respect to the probability law induced
by $\tilde{W}(t)$. Hence, with probability 1, the sample paths of
$\rho(\cdot)$ again lie in $\bar{\mathbb{C}}[0,T]$. Then, by similar
arguments as in the case with $\bm{h}=(0,0)$, but now using (\ref{eq:convergence=000020of=000020rho_n=000020general}),
we conclude 
\begin{equation}
(\tau_{n,T},\delta_{n,T})\xrightarrow[P_{n,\bm{h}}]{d}(\tau_{T}^{*},\delta_{T}^{*}).\label{eq:joint=000020convergence=000020of=000020tau,=000020delta}
\end{equation}

\subsubsection*{Step 3 (Convergence of $V_{n}(\bm{d}_{n,T},\bm{h})$)}

From (\ref{eq:evolution=000020of=000020rho}) and the discussion in
Section \ref{subsec:Intuition-behind-Theorem}, it is clear that the
distribution of $\rho(t)$ is the same as that of $\sigma_{1}^{-1}x_{1}(t)-\sigma_{0}^{-1}x_{0}(t)$
in the diffusion regime. Thus, the joint distribution, $\mathbb{P}$,
of $(\tau_{T}^{*},\delta_{T}^{*})$, defined in Step 2, is the same
as the joint distribution of 
\[
\left(\tau_{T}^{*}:=\tau^{*}\wedge T,\delta_{T}^{*}:=\mathbb{I}\left\{ \frac{x_{1}(\tau^{*}\wedge T)}{\sigma_{1}}-\frac{x_{0}(\tau^{*}\wedge T)}{\sigma_{0}}\ge0\right\} \right)
\]
in the diffusion regime, when the optimal sampling rule $\pi^{*}$
is used. Therefore, defining $\bm{d}_{T}^{*}\equiv(\pi^{*},\tau_{T}^{*},\delta_{T}^{*})$
and $\mathbb{E}[\cdot]$ to be the expectation under $\mathbb{P}$,
we obtain 
\[
V(\bm{d}_{T}^{*},\mu(\bm{h}))=\mathbb{E}\left[\Delta\mu(\bm{h})\delta_{T}^{*}+c\tau_{T}^{*}\right],
\]
where $V(\bm{d},\bm{\mu})$ denotes the frequentist regret of $\bm{d}$
in the diffusion regime. Now, recall that by the definitions stated
early on in this proof,
\[
V_{n}(\bm{d}_{n,T},\bm{h})=\mathbb{E}_{n,\bm{h}}\left[\sqrt{n}\Delta_{n}\mu(\bm{h})\delta_{n,T}+c\tau_{n,T}\right].
\]
Since $\delta_{n},\tau_{n}$ are bounded and $\sqrt{n}\Delta_{n}\mu(\bm{h})\to\Delta\mu(\bm{h})$
by Assumption 1(iii), it follows from (\ref{eq:joint=000020convergence=000020of=000020tau,=000020delta})
that for each $\bm{h}$, 
\begin{equation}
\lim_{n\to\infty}V_{n}(\bm{d}_{n,T},\bm{h})=V(\bm{d}_{T}^{*},\mu(\bm{h})).\label{eq:pf:Thm2=000020pointwise=000020convergence}
\end{equation}

For any given $\bm{h}$ and $\epsilon>0$, a dominated convergence
argument as in Step 4 of the proof of Theorem \ref{Thm:=000020Parametric=000020families=000020lower=000020bound}
shows that there exists $\bar{T}_{\bm{h}}$ large enough such that
\begin{equation}
V(\bm{d}_{T}^{*},\mu(\bm{h}))\le V(\bm{d}^{*},\mu(\bm{h}))+\epsilon\label{eq:pf:Them=0000203:=000020bound=000020on=000020time}
\end{equation}
for all $T\ge\bar{T}_{h}$. Fix a finite subset $\mathcal{J}$ of
$\mathbb{R}$ and define $\bar{T}_{\mathcal{J}}=\sup_{\bm{h}\in\mathcal{J}}T_{\bm{h}}.$
Then, (\ref{eq:pf:Thm2=000020pointwise=000020convergence}) and (\ref{eq:pf:Them=0000203:=000020bound=000020on=000020time})
imply 
\[
\liminf_{n\to\infty}\sup_{\bm{h}\in\mathcal{J}}V_{n}(\bm{d}_{n,T},\bm{h})\le\sup_{\bm{h}\in\mathcal{J}}V(\bm{d}_{T}^{*},\mu(\bm{h}))\le\sup_{\bm{h}\in\mathcal{J}}V(\bm{d}^{*},\mu(\bm{h}))+\epsilon,
\]
for all $T\ge\bar{T}_{\mathcal{J}}$. Since the above is true for
any $\mathcal{J}$ and $\epsilon>0$, 
\begin{align*}
\sup_{\mathcal{J}}\lim_{T\to\infty}\liminf_{n\to\infty}\sup_{\bm{h}\in\mathcal{J}}V_{n}(\bm{d}_{n,T},\bm{h}) & \le\sup_{\mathcal{J}}\sup_{\bm{h}\in\mathcal{J}}V(\bm{d}^{*},\mu(\bm{h}))\\
 & \le\sup_{\bm{\mu}}V(\bm{d}^{*},\bm{\mu})=V^{*}.
\end{align*}
The inequality can be made an equality due to Theorem \ref{Thm:=000020Parametric=000020families=000020lower=000020bound}.
We have thereby proved Theorem \ref{Thm:=000020Parametric=000020-=000020attaining=000020the=000020bound}.
\end{document}